\documentclass[journal=jpcbfk,manuscript=article]{achemso}

\usepackage[version=3]{mhchem} 
\usepackage{graphicx,amssymb,amstext,amsmath} 
\usepackage{color}
\usepackage{wasysym}
\usepackage{bm}

\SectionNumbersOn

\author{Richard Matthews}

\affiliation[]
{Faculty of Physics, University of Vienna, Boltzmanngasse 5, A-1090 Vienna, Austria}

\email{richard.matthews@univie.ac.at}

\author{Christos N. Likos}

\affiliation[]
{Faculty of Physics, University of Vienna, Boltzmanngasse 5, A-1090 Vienna, Austria}

\title[]
  {Dynamics of Self-Assembly of Model Viral Capsids in the Presence of a Fluctuating Membrane \footnote{This document is the Accepted Manuscript version of a work that will appear in final form in The Journal of Physical Chemistry B, \copyright~American Chemical Society after peer review and technical editing by the publisher.}}

\begin{document}

\begin{abstract}
A coarse-grained computational model is used to investigate the effect of a fluctuating fluid membrane on the dynamics of patchy-particle assembly into virus capsid-like cores. Results from simulations for a broad range of parameters are presented, showing the effect of varying interaction strength, membrane stiffness and membrane viscosity. Furthermore, the effect of hydrodynamic interactions is investigated. Attraction to a membrane may promote assembly, including for sub-unit interaction strengths for which it does not occur in the bulk, and may also decrease single-core assembly time. The membrane budding rate is strongly increased by hydrodynamic interactions. The membrane deformation rate is important in determining the finite-time yield. Higher rates may decrease the entropic penalty for assembly and help guide sub-units towards each other but may also block partial cores from being completed. For increasing sub-unit interaction strength, three regimes with different effects of the membrane are identified. 
\end{abstract}


\pagebreak


\section{Introduction}
\label{sec:intro}

The formation of the protein shell of viruses has, due to its relative simplicity and importance in many diseases, become one of the most well-studied examples of self-assembly~\cite{hagan2013}. Although viruses are typically assembled within the cells of their host, the process may also be triggered in a bulk solution of viral proteins by changing the pH~\cite{fraenkel}. Such experiments have stimulated the application of simple computational models~\cite{rapaport, nguyen, hagan, wilber, johnston,hagan2013} to help understand assembly processes.

Whilst much modeling has focussed on the formation of virus capsids in the bulk, in recent work investigating the growth of viral shells around their genome, the assembly of simple sub-units attracted to a flexible polymer was simulated~\cite{elrad2010,mahalik2012}. Interaction with the polymer was found to allow assembly for parameters for which it would otherwise not occur. Encapsulation of spherical nano-particles has also been considered both in experiment~\cite{sun2007} and in simulation~\cite{elrad2008,williamson2011}. Experimentally, it was demonstrated that shells resembling different types of viral particles could be assembled by varying the nano-particle diameter.

Beyond interactions with an encapsulated genome, there is also much evidence that membranes play an important role in assembly for many viruses~\cite{gelderblom,ono,miyanari,shavinskaya,forsell,ng,simon,siegel,bravo}. In a recent publication~\cite{matthews2012}, we presented results on the effect of fluctuating membranes on the equilibrium of a system of self-assembling patchy colloids, designed to assemble viral core-like structures, from Monte Carlo (MC) simulations~\cite{frenkel}. We found a non-monotonic dependence of the promotion of assembly on membrane stiffness, as well as the formation of membrane buds. It is of course true that such effects would be observable in an analogous experimental system after sufficient time and to be expected that they will influence the products of dynamical assembly. However, on relevant timescales, self-assembly processes may not reach equilibrium and the products may be affected, for example, by kinetic traps~\cite{zlotnick2000,hagan2013}. It is therefore of foremost interest to consider simulations with realistic dynamics. Key dynamical features that we capture in our simulations are the viscosity of the membrane and hydrodynamic interactions, the inclusion of which may alter dynamics both quantitatively and qualitatively~\cite{kikuchi2002}.

Two key factors in the present work are attractions to the fluctuating membrane and hydrodynamic interactions. Previous computational studies have studied the effect of each of these individually on the clusters formed by isotropic spherical colloids. Hydrodynamic interactions were found to change both the size and shape of clusters~\cite{whitmer2011}, whilst attraction to a membrane was found to induce the formation of linear chains on the surface~\cite{saric2012}. Further, attractions of particles to a membrane surface may cause the formation of buds~\cite{reynwar2007,zhang2008,matthews2012} or tube-like structures~\cite{saric2012tube,bahrami2012}. 

Here, as a simple model to gain insight into the effect of membranes on the dynamics of self-assembly, we consider primarily the same, patchy-particle, sub-units~\cite{wilber,bianchi}, which may assemble twelve-component cores, as in our previous work~\cite{matthews2012}, and simulate their assembly using a dynamically realistic method. As previously, our sub-units are coupled to a membrane modeled using particles bonded to form a triangulated surface~\cite{gompper1996,noguchi2005}. The target core structure has icosahedral symmetry, similar to many viruses, although in reality enveloped viruses are larger. The remainder of the paper is organized as follows. In section~\ref{sec:models} we describe our simulation models and in section~\ref{sec:equ_results} present results from MC simulations on the equilibrium of the system. We then move on to dynamical simulations, describing simulation methods in section~\ref{sec:dyn_meth}. We present results for the twelve-component cores in section~\ref{sec:dyn_results} and compare them to some results for some other cores in section~\ref{sec:other_target}. Finally, we conclude in section~\ref{sec:conc}.

\section{Simulation Models}
\label{sec:models}

\begin{figure*}[ht!]
\begin{center}

\includegraphics[scale=0.6]{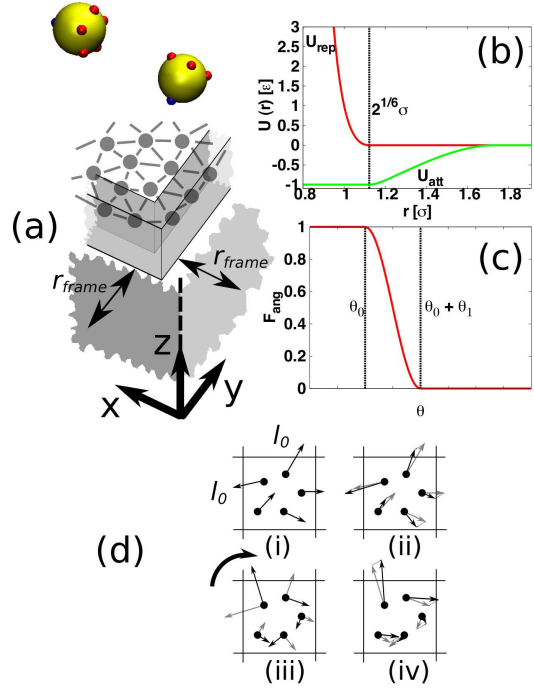}

\caption{\label{fig:set_up} (a) Simulation set-up. Sub-units, which are all identical, are rendered in yellow, with positions, but not extents, of patches for interactions with other sub-units in red. Positions of patches for interactions with the membrane particles are in blue. The membrane is modeled as a triangulated surface of bonded particles. The particles forming the surface edge are confined to a frame region, which is located at a distance $r_{frame}$ from the periodic boundaries. In simulations with hydrodynamics, a stochastic rotation dynamics (SRD) solvent composed of point particles is included. Interactions between SRD particles are effected by first dividing the entire system into a grid of cells of side $l_0$. (b) The radial part of the inter-sub-unit or sub-unit-membrane potential, $U(r)$, with a well-depth $\epsilon$ is split into attractive (green) and repulsive parts (red). (c) The attractive part is multiplied by factors of the form $F_{ang}(\theta)$, where $\theta$ is an angle that depends on the relative orientation of the interacting particles. (d) Sketch of momentum transfer between SRD particles in a cell: (i) Only particles within one cell interact. (ii) Velocities are subtracted from all particles such that the centre of mass velocity is 0. (iii) All velocities are rotated, as signified by the heavy arrow, around a random axis, by a given angle. (iv) The subtracted velocities are added back on so that total momentum is conserved.}
\end{center}
\end{figure*}

Rather than only considering enough sub-units to form just one target structure as in our previous work~\cite{matthews2012}, we now simulate $180$, allowing a maximum of 15 complete cores to be assembled. Whilst it is expected that in experimental and biological situations it is also likely that a larger number of sub-units will be available than required for one complete structure, this choice was additionally made for computational efficiency, so that, on a feasible timescale, although assembly of all possible cores may not occur, some complete cores will form. We simulate a membrane composed of 1156 particles. The simulation set-up is sketched in Fig.~\ref{fig:set_up}(a).

The interactions between sub-units, $ss$, and between sub-units and membrane particles, $ms$, are identical to those used in our previous work~\cite{matthews2012} but we describe the important features again here. The potentials are based on a Lennard-Jones form. As shown in Eq.~(\ref{eq:LJ_pot}), the potential is split into attractive, $U_{att}$, and repulsive, $U_{rep}$, parts. The interaction of two particles, $i$ and $j$, separated by $\boldsymbol{r}_{ij}$ ($i \neq j$), with orientations $\boldsymbol{\Omega}_i$ and $\boldsymbol{\Omega}_j$, either both sub-units or a sub-unit and a membrane particle, is given by
\begin{equation}
U_{ij}(\boldsymbol{r}_{ij}, \boldsymbol{\Omega}_i,\boldsymbol{\Omega}_j)  =\gamma_{area} \left[U_{rep}(r_{ij}) + \gamma_{att} \gamma_{orient} U_{att}(r_{ij}) \right],
\label{eq:LJ_pot}
\end{equation}
where the forms of $U_{att}$ and $U_{rep}$ are shown in Fig.~\ref{fig:set_up}(b). $\gamma_{area}$, $\gamma_{att}$ and $\gamma_{orient}$ are dimensionless factors that take different forms for $ss$ and $ms$ interactions. For $ss$-interactions, $\gamma_{area} = \gamma_{att} = 1$ and, as depicted in Fig.~\ref{fig:set_up}(a), there are 5 patches on each sub-unit, which are arranged symmetrically around a single $ms$ patch. The minimum of $U_{att}$ is set to $-\epsilon_{ss}$. $\gamma_{orient}$ is used to control the patch width, and it has the form of a product of three functions of the form shown in Fig.~\ref{fig:set_up}(c), see also the Supporting Information. For the first two factors, the argument is the angle between the interacting patches and the centre-to-centre vector, $\boldsymbol{r}_{ij}$. The parameters for determining patch width, see Fig.~\ref{fig:set_up}(c), are set to $\theta_0 = \theta_1 = 0.2$. In contrast, for the third factor, the argument is the angle between the projections of the membrane patch onto the plane perpendicular to $\boldsymbol{r}_{ij}$ and $\theta_0 = \theta_1 = 0.4$. The third factor represents the torsional stiffness of protein interactions~\cite{wilber}.

For $ms$-interactions, the minimum of $U_{att}$ is set to $-\epsilon_{ms}$. In these interactions, only the sub-units are patchy, having one patch. Parameters for the one orientational function composing $\gamma_{orient}$, see Fig.~\ref{fig:set_up}(c), are $\theta_0$ = $\pi/4$ and $\theta_1 = 0.2$, and there is no penalty for sub-units rotating around $\boldsymbol{r}_{ij}$. Since, typically, assembling proteins will only be able to access one side of a membrane, we choose to make only one side of the membrane in our simulations attractive to sub-units~\cite{matthews2013}. This is achieved by setting $\gamma_{att} = 1$ if a sub-unit interacts with the ``upper'' side and $\gamma_{att} = 0$ if it interacts with the ``lower'' side. The $\gamma_{area}$ factor is proportional to the area of surface that surrounds the interacting membrane particle. The length scale for $ss$-interactions is chosen as $\sigma_{ss} = 2.5l_0$ and the length scale for $ms$ interactions is $\sigma_{ms} = 1.75l_0$. For the exact functional forms used in the $ss$ and $ms$ interactions, see the Supporting Information. 

The membrane is modeled as in ref.~\cite{matthews2013} but we describe the key features again here. As depicted in Fig.~\ref{fig:set_up}(a), the membrane is composed of particles bonded to form a triangulated surface. To include membrane fluidity, MC moves that flip bonds between different particles are included.~\cite{noguchi2005}. The typical separation between bonded membrane particles is $l_0$, maintained by a potential that has a flat central region but diverges at $0.67l_0$ and $1.33l_0$, see Supporting Information. We perform simulations in a box of size $45 l_0 \times 45 l_0 \times 45 l_0$, giving a sub-unit number density within the range for which yield was found to weakly depend on concentration~\cite{wilber}. As in our previous work~\cite{matthews2012}, we consider a range of $\epsilon_{ss}$ that, at equilibrium in the bulk, covers the crossover to complete assembly of all cores. Although approximately centered around the same $\epsilon_{ss}$ value, for the larger number of cores considered here, the crossover is broader~\cite{ouldridge2010} and so a wider range of $\epsilon_{ss}$ is used. The same range of $\epsilon_{ms}$ as in ref.~\cite{matthews2012} is considered, chosen to cover the crossover from freely diffusing to membrane-bound structures. The stiffness of our membrane is controlled by a parameter $\lambda_b$, through a potential, $U_{bend} = \lambda_b (1 - \boldsymbol{n}_i \cdot \boldsymbol{n}_j)$, applied to all pairs of neighboring triangles in the surface, where $ \boldsymbol{n}_i$ and $ \boldsymbol{n}_j$ are the unit normal vectors of the triangles. We simulate using the three middle values from our previous work~\cite{matthews2012}, $\lambda_b = \sqrt{3} k_BT$, $2\sqrt{3} k_BT$ and $4\sqrt{3} k_BT$: at equilibrium, this covers the crossover from cores being able to cause budding of the membrane to them not being able to. As discussed in our previous work~\cite{matthews2012}, this range of bending stiffness is at the lower end of that expected for biological membranes. Given that in viral budding~\cite{lerner1993} intrinsic curvature is expected to be important, which is neglected in our model, the bending stiffness in our simulation is most relevant in terms of the cost of deformation.

Although our focus is on dynamical simulation, we first investigate the equilibrium of the system for comparison. For this purpose, we use MC simulations, employing a similar approach as in our previous work~\cite{matthews2012}. On the other hand, for molecular dynamics (MD) simulations, we include hydrodynamic interactions using a stochastic rotation dynamics (SRD) solvent~\cite{gompper2009}, a coarse-grained method in which the fluid is represented by point particles. SRD particle interactions are effected by dividing the system into a grid of cells, of side $l_0$, at regular intervals and exchanging momentum by a rotation through a certain angle of velocities relative to the cell centre of mass velocity. This procedure is shown schematically in Fig.~\ref{fig:set_up}(d). To understand the influence of hydrodynamic interactions, we also simulate using a method that neglects them, Langevin dynamics (LD), in which the effect of the solvent is represented by uncorrelated random, as well as drag, forces~\cite{dunweg2003,ladd2009}.

To simulate a tensionless membrane, rather than box rescaling~\cite{matthews2012}, we use a new membrane boundary condition, recently introduced by us~\cite{matthews2013}, which is compatible with SRD. The edge of the membrane is attached to a square frame, with sides positioned at a distance $r_{frame}$ into the simulation box, as depicted in Fig.~\ref{fig:set_up}(a). For those triangles in the surface that have a side that forms part of the membrane edge, a bending potential of the same form as that between neighboring triangles is applied, except that  the unit normal of the triangle is compared to a unit normal to the frame-plane. The distance $r_{frame}$ may increase and decrease during the simulation. To allow for deformation, the number of membrane particles bonded to the frame may also vary, with corresponding changes to the number of bonds in the bulk of the surface, $N_{b-bulk}$. For more details of the membrane boundary condition, see the appendix of ref.~\cite{matthews2013}, the functional form of the confining potential is also given in the Supporting Information. For consistency, this approach is also used in MC and LD simulations, in which, of course, the solvent is absent.

\section{Results from equilibrium simulations}
\label{sec:equ_results}

\begin{figure*}[ht]
\begin{center}

\includegraphics[scale=0.4]{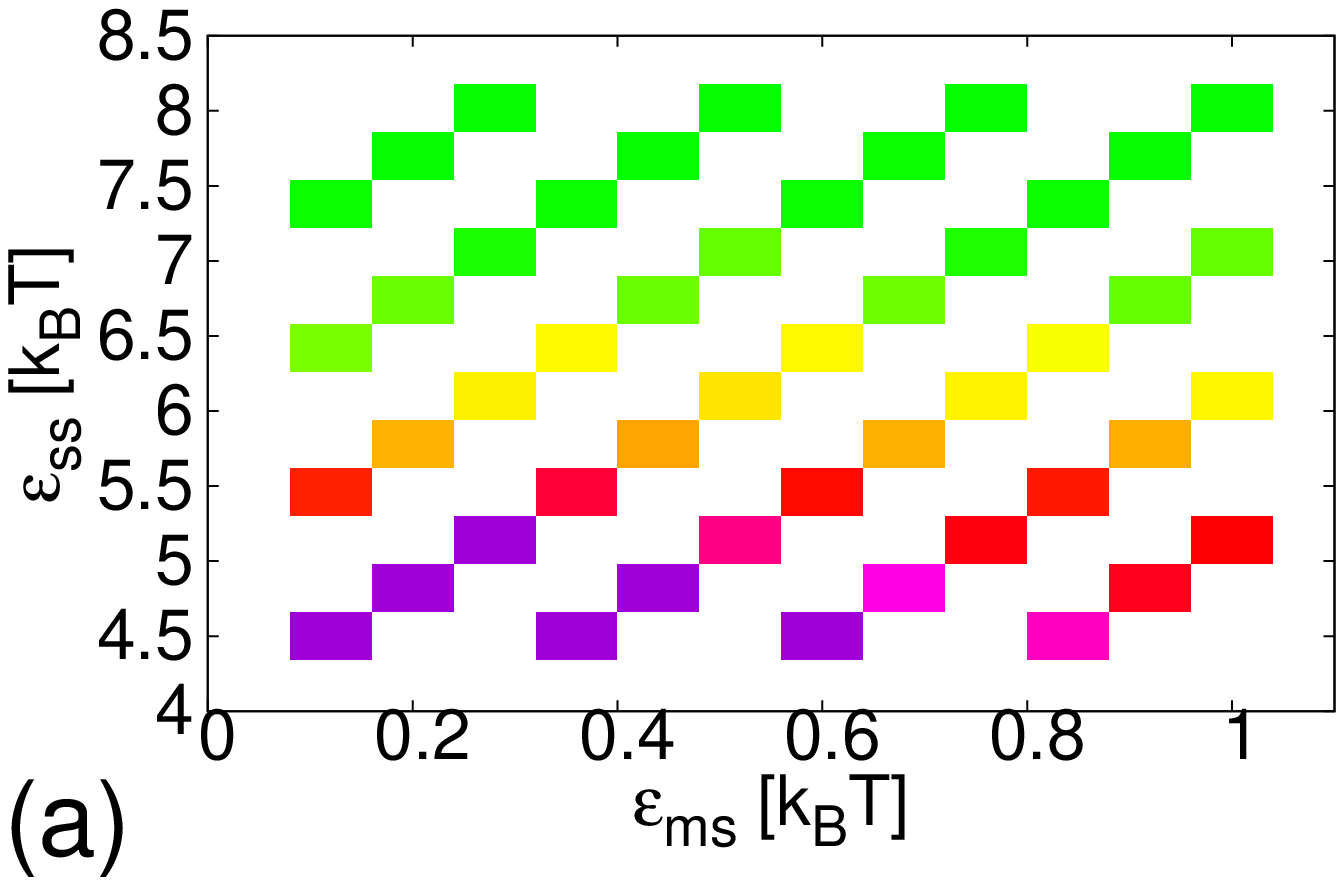}
\includegraphics[scale=0.4]{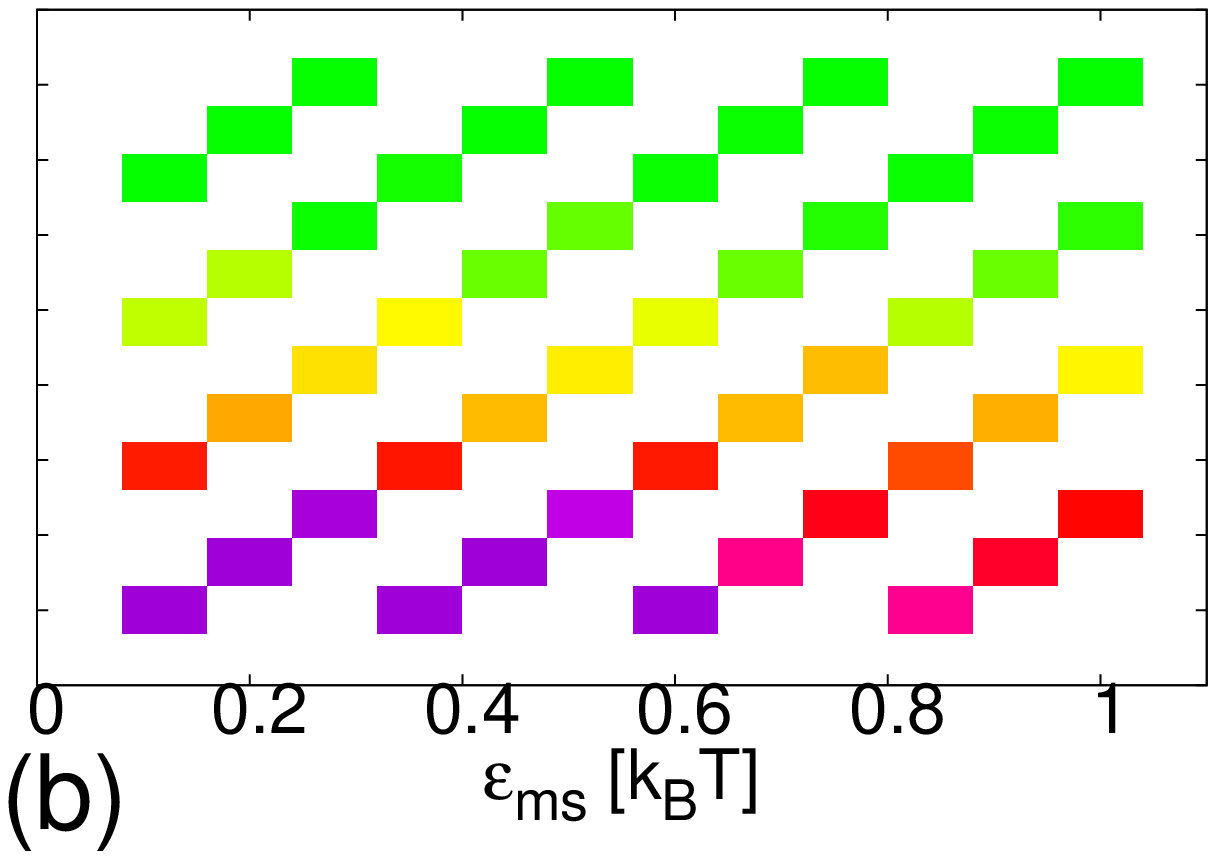}
\includegraphics[scale=0.4]{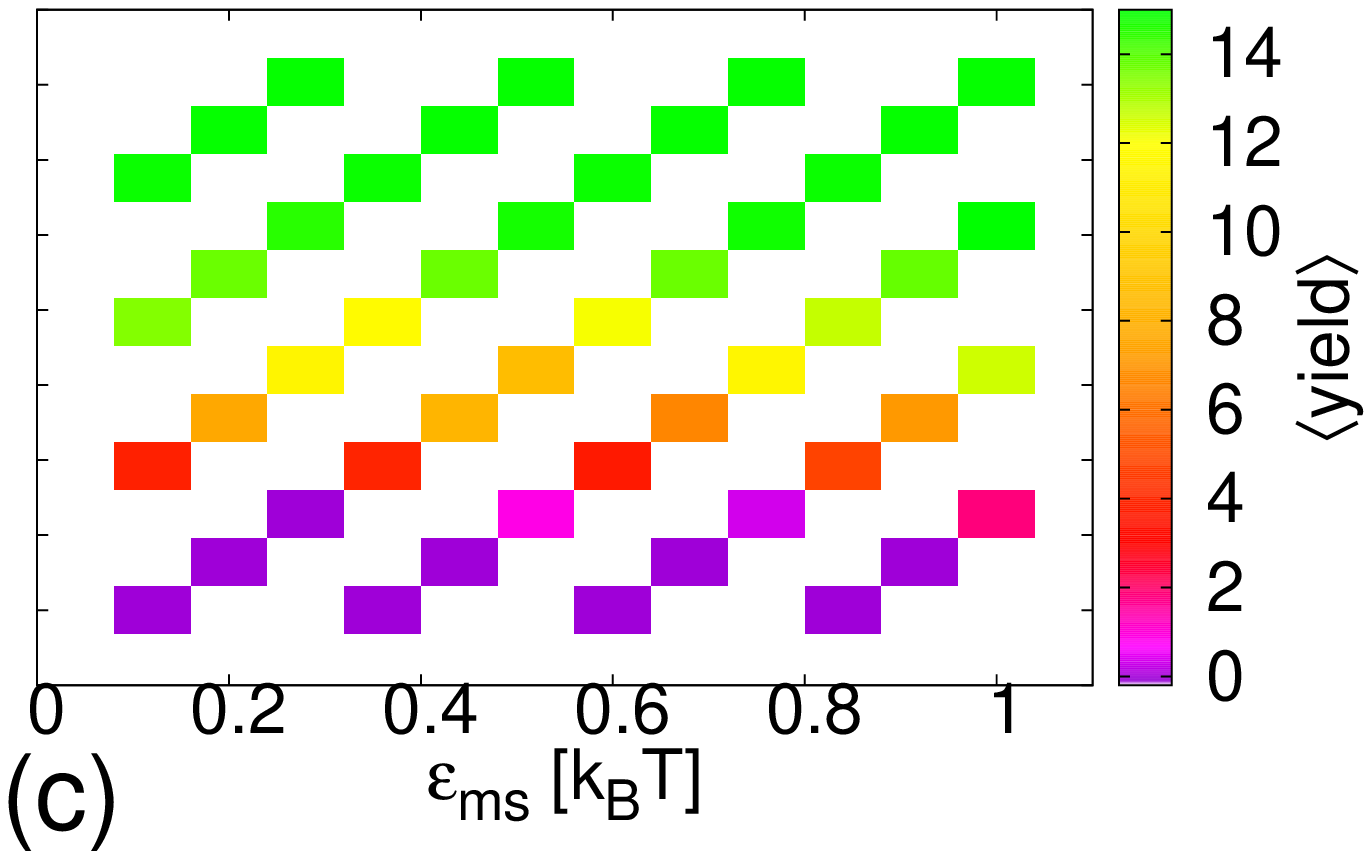}

\caption{\label{fig:MC_yield} Results from MC simulations. Average yield of complete cores, $\left<yield\right>$, as a function of sub-unit-membrane interaction strength, $\epsilon_{ms}$, and inter-sub-unit interaction strength, $\epsilon_{ss}$, for different membrane stiffnesses, $\lambda_b$: (a) $\lambda_b = \sqrt{3}k_BT$;  (b) $\lambda_b = 2\sqrt{3}k_BT$; (c) $\lambda_b = 4\sqrt{3}k_BT$.}
\end{center}
\end{figure*}

We first present, in Fig.~\ref{fig:MC_yield}, results from MC simulations on the yield of complete cores, defined to be a cluster of 12 bonded sub-units, each unit making 5 bonds to other cluster members. Two sub-units are defined to be bonded if their interaction energy is $< -0.25\epsilon_{ss}$. Interaction strengths for different simulations lie on a grid from $0.12$ to $k_BT$ in spacings of $0.08k_BT$ for $\epsilon_{ms}$ and from $4.5$ to $8.02 k_BT$ in spacings of $0.32k_BT$ for $\epsilon_{ss}$. Systems at different parameters were run in parallel using Multicanonical Parallel Tempering~\cite{faller}. For each data point in Fig.~\ref{fig:MC_yield}, approximately $4\times10^9$ attempted MC moves were performed, including about $4\times10^4$ Hybrid MC moves~\cite{mehlig}, as well as Aggregate Volume Bias moves~\cite{chen_AVBMC}. These were both found to significantly speed up relaxation. The largest error for a single data point was estimated to be about 0.6. Similarly to our results with one core~\cite{matthews2012}, for $\lambda_b = \sqrt{3} k_BT$ and $2\sqrt{3} k_BT$ at higher $\epsilon_{ms}$, the assembly of the cores causes the membrane to form buds, although these now generally contain multiple cores. For $\lambda_b = 4\sqrt{3} k_BT$, budding did not occur. Again as for single cores, for high $\epsilon_{ms}$, assembly occurs for lower values of $\epsilon_{ss}$: the membrane promotes assembly. Here, membrane-dependent, low $\epsilon_{ss}$ assembly does not occur to the same extent as for high $\epsilon_{ss}$ because, due to steric repulsion, only a fraction of the cores may interact with the membrane at once, typically about 4 cores in the case where a bud is formed.  Whereas for one core, the range over which promotion occurred was clearly largest for $\lambda_b = \sqrt{3} k_BT$, here the results for  $\lambda_b = 2\sqrt{3} k_BT$ are very similar. This may be because multiple cores together effectively form a larger object deforming the membrane.

\section{Dynamical simulation methods}
\label{sec:dyn_meth}

We next give details of our dynamical simulation methods. The SRD particles have mass $m$ and number density per cell $\gamma = 5$. We define our unit of time, $t_0 = l_0\sqrt{m/k_BT}$. Collisions are performed every $\Delta t_{coll} = 10^{-1}t_0$ and we use an SRD rotation angle of $\frac{\pi}{2}$, giving a fluid viscosity of $\eta_f = 2.5 m/l_0 t_0 $~\cite{kikuchi2003}. We apply a SRD-cell level thermostat that conserves momentum to maintain the temperature~\cite{gompper2009}. Membrane particles are coupled to the SRD solvent by including them in the collision step~\cite{gompper2009}. There will typically be about one membrane particle per SRD cell and we set their mass to $\gamma m$, giving a short-time friction coefficient $\zeta_{mem} =  15.8(m / t_0)$~\cite{kikuchi2003}.

Unlike membrane particles, sub-units have rotational degrees of freedom and so are coupled to the SRD solvent using bounce-back boundary conditions~\cite{whitmer2010}. For their interactions with the fluid, sub-units are treated as solid spherical particles of radius $a = l_0$, having mass $M = \frac{4}{3}\pi a^{3} m \gamma$ and moment of inertia $I = \frac{2}{5}Ma^{2}$. Every $\Delta t_{bound} = 10^{-2}t_0$, the SRD particles are checked. If an overlap with a sub-unit is detected, then the SRD particle with velocity $\boldsymbol{u}$ is first moved by $-\frac{1}{2}\Delta t_{bound} \boldsymbol{u}$ and then shifted radially to the edge of the sub-unit, $\boldsymbol{r}$ from the centre, where $\left| \boldsymbol{r} \right | = a$. This scheme is based on the fact that for SRD particles the average crossing of the sub-unit boundary is halfway through a time step. It was found to function well in previous work~\cite{padding2005}. A bounce-back collision is then performed: the radial, $\boldsymbol{u}_{\perp}$, and tangential, $\boldsymbol{u}_{\parallel}$, components of $\boldsymbol{u}$ are updated according to
\begin{eqnarray}
\boldsymbol{u}^{new}_{\perp} &=& \left(1 - A \right)\boldsymbol{u}^{old}_{\perp}  + A \boldsymbol{V}_{\perp} \nonumber \\
\boldsymbol{u}^{new}_{\parallel} &=&  -\frac{1 - B}{1 + B} \boldsymbol{u}^{old}_{\parallel} + \frac{2}{1 + B}\boldsymbol{V}_{\parallel}.
\label{eq:bounce_back}
\end{eqnarray}
Here, $A= \frac{2M}{(m+M)}$, $B = \frac{7m}{2M}$ and the surface velocity $\boldsymbol{V} = \boldsymbol{v} + \boldsymbol{\omega} \times \boldsymbol{r}$, where $\boldsymbol{v}$ is the centre of mass velocity of the sub-unit and $\boldsymbol{\omega}$ is its angular velocity around an axis that passes through the centre of mass. Eq.~\ref{eq:bounce_back} is valid for $I = \frac{2}{5} M a^2$. After all overlapping SRD particles have been rebounded, corresponding changes to the sub-unit velocity and angular velocity, $\Delta \boldsymbol{v} = \frac{m}{M}\sum\limits_{i}\left(\boldsymbol{u}_i^{old} - \boldsymbol{u}_i^{new}\right)$ and $\Delta \boldsymbol{\omega} = \frac{m}{I}\sum\limits_{i}\boldsymbol{r}_i \times\left(\boldsymbol{u}_i^{old} - \boldsymbol{u}_i^{new}\right)$, where $i$ indexes the different rebounded particles, are applied so that momentum and energy are conserved. If $M \gg m$, SRD particle velocities relative to the surface are completely reversed; for our parameters $M \approx 20m$.

Overlapping of embedded particles in an SRD fluid may lead to a spurious depletion attraction~\cite{padding2006}. In fact, even if particles are prevented from overlapping, the bounce-back scheme may need to be iterated due to the possibility of a fluid particle interacting with more than one solute particle within $\Delta t_{bound}$. We avoid these issues by choosing the excluded volume length for sub-unit interactions, $\sigma_{ss} = 2.5l_0$, so that the typical closest approach of two sub-unit fluid surfaces $\approx 0.5 l_0$ is much greater than the typical displacement of a fluid particle $\approx 10^{-2} l_0$.

Bounce-back interactions between SRD particles and embedded colloids lead to spurious slip at the colloid surface. Methods exist to ameliorate this by the introduction of virtual particles but, for mobile colloids, this was found to lead to deviations from expected thermal distributions~\cite{whitmer2010}. In our simulations, the concern is moot anyway, because of the discrepancy between the radii chosen for inter-sub-unit and sub-unit-fluid interactions. Effectively, there is a slip-velocity at the sub-unit surface, as defined by its interactions with other sub-units, which has contributions from these two different sources. Given that the sub-units are typically representing protein complexes, which are rough on length-scales up to many solvent molecules~\cite{pettit1999}, rather than smooth colloids, this is reasonable.

For bounce-back boundaries, the short-time friction coefficients for the sub-units may be calculated using a modified Enskog theory~\cite{whitmer2010}. For our parameter choice, this gives coefficients of $\zeta_{v} = 62.0 (m / t_0)$ and $\zeta_{\omega} = 73.3 (m l_0^2 / t_0)$, for linear and angular velocities respectively. Comparing the corresponding correlation times, $M/\zeta_{v}$ and  $I/\zeta_{\omega}$, to typical thermal velocities, we obtain values of $0.07 l_0$ and $0.04$ for the typical length and angular displacements over which the sub-unit motion is correlated. These are smaller than the typical separation of sub-units and patches, given by $\sigma_{ss} = 2.5 l_0$ and $\approx 0.4$ respectively, so that at the scale that assembly occurs on, sub-unit motion is diffusive. Similarly, the length scale over which membrane particle motion is correlated is $0.1l_0$.

To obtain parameters for simulations without hydrodynamic interactions, we simulated single sub-units and membrane particles in a box of the same size as that used for assembly with an SRD solvent. The friction coefficients extracted were lower than the short-time values due to long-time hydrodynamic contributions. These friction coefficients were input to LD simulations. In this way, the hydrodynamic contribution to the self-diffusion coefficients is included but hydrodynamic interactions between different particles are neglected. An alternative approach to simulating without hydrodynamics is to use an SRD fluid and randomize particle velocities at every step. For colloids, however, this has been found to introduce an unphysical caging effect~\cite{belushkin2012}.

Membrane fluidity is included by performing a certain number of attempts to flip bonds between neighboring pairs of membrane particles~\cite{noguchi2005} every $10^{-1}t_0$. The membrane viscosity is set by the level of attempted bond-flips and we consider three different rates: $N_{b-bulk}$, $10^{-1}N_{b-bulk}$ and $10^{-2}N_{b-bulk}$ attempted bond-flips per $10^{-1}t_0$, where $N_{b-bulk}$ is the number of bonds in the bulk of the membrane and the resulting numbers are rounded to integers. By considering Poiseuille flows in two-dimensional membranes~\cite{noguchi2005}, the corresponding membrane viscosity, $\eta_m$, may be estimated. For the highest rate of flips the value is estimated to be $35.1\pm0.1 m/t_0$~\cite{matthews2013}, whereas for the lower rates we estimate $133.3\pm0.6 m/t_0$ and $1190\pm60 m/t_0$ respectively. For a lipid bilayer in water, the ratio of membrane to fluid viscosities, $l_{\eta}$, is typically around $1 - 10 \mu$m~\cite{petrov2008}. In our simulations the solvent viscosity $\eta_f = 2.5 m/l_0 t_0$ so that, if our sub-units represent capsomers with a size on the order of $10$nm~\cite{baker1999}, then the ratio of their hydrodynamic radius to $l_{\eta}$ is around the expected range.

\section{Results from dynamical simulations}
\label{sec:dyn_results}

We next present results from our dynamical simulations. Interaction strengths for dynamical simulations were chosen to coincide with those for MC simulations, although fewer were considered due to higher computational costs. A closer spacing between the highest interaction strengths was chosen as it was expected that the most interesting results would be found here. All averages are taken over at least five independent runs and in some cases over ten. We consider the same values of $\lambda_b$, and also $\epsilon_{ms}$ and $\epsilon_{ss}$ in the same range, as for the equilibrium MC simulations. We simulate primarily using SRD but, for $\lambda_b = \sqrt{3} k_BT$ and $2 \sqrt{3} k_BT$, we also simulate using LD for comparison, to gain insight into the importance of hydrodynamic interactions. LD simulations were essentially identical to the SRD ones expect that, rather than having regular interactions with an explicit fluid, sub-units and membrane particles were, at each MD integration step, subject to random and friction forces~\cite{dunweg2003,ladd2009}. The system was initially simulated for either $8\times 10^3 t_0$ with SRD, or $2\times 10^4 t_0$ with LD, without attractive interactions. These times were chosen as being sufficient to allow membrane relaxation. Subsequently, attractions were switched on and the system was simulated for a further $5\times 10^4 t_0$ to gather results. In contrast to the MC results, for the stiffest membrane, $\lambda_b = 4\sqrt{3}k_BT$, with the highest membrane-sub-unit interaction strength only, $\epsilon_{ms} = k_BT$, in some, though not all runs, budding occurred.

Were it possible to run the dynamical simulations indefinitely, it is expected that results would eventually converge to those found for the equilibrium simulations. However, as the simulation progresses, further assembly becomes increasing slow as the supply of free sub-units is depleted and eventually relies on rearrangement of sub-units between partially formed structures, possibly moving into or out of a membrane bud. It is thus necessary to choose a finite simulation time shorter than that required for complete assembly and inevitably the results obtained will depend on it. For our chosen simulation time, the maximum yield observed in any simulation is $\approx 50\%$ of the possible maximum. It is nonetheless sufficient for the effect of the membrane on assembly to be apparent. However, the finite time chosen should be borne in mind when considering the results.

\begin{figure*}[ht]
\begin{center}

\includegraphics[scale=0.27]{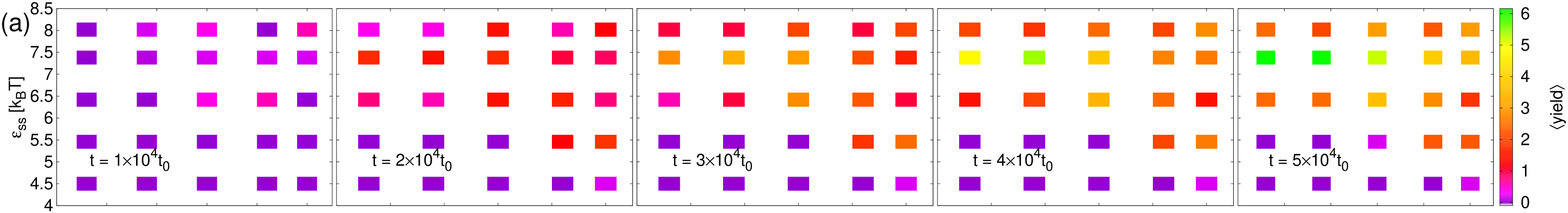}
\includegraphics[scale=0.27]{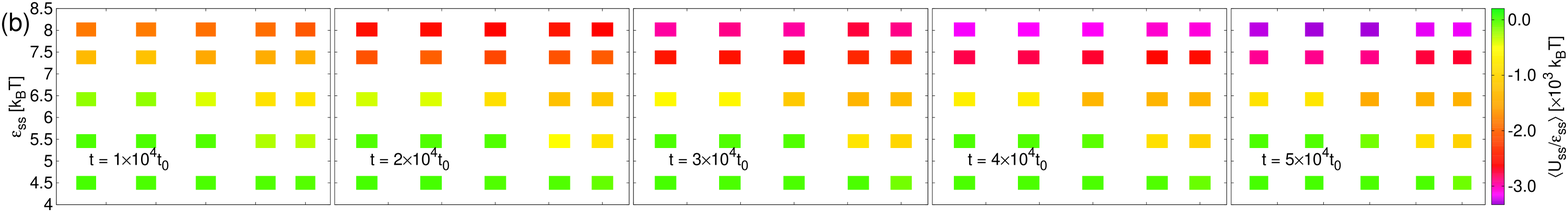}
\includegraphics[scale=0.27]{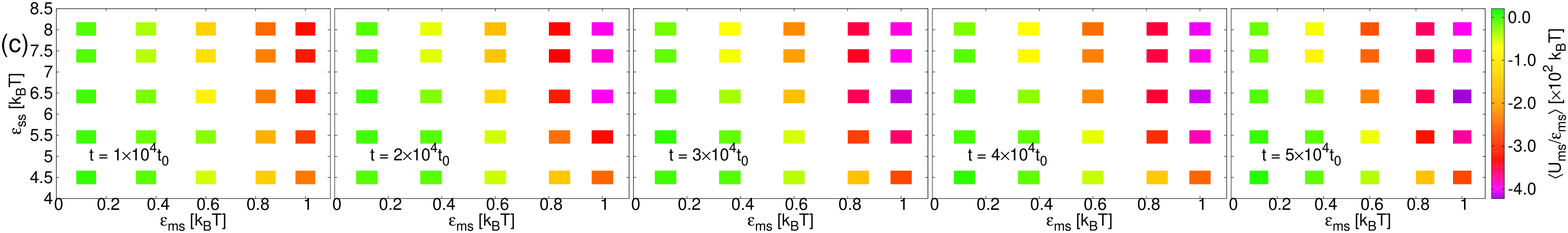}

\caption{\label{fig:yield_time_visc_0.1_kap_2} Plots as a function of sub-unit-membrane interaction strength, $\epsilon_{ms}$, and inter-sub-unit interaction strength, $\epsilon_{ss}$ at different times, $t$, increasing from left to right in intervals of $1\times10^4 t_0$, for membrane stiffness, $\lambda_b = 2\sqrt{3} k_BT$ and membrane viscosity $\eta_m = 133.3 m/t_0$, from SRD simulations. (a) The average yield of complete cores, $\left<yield\right>$. (b) The average total interaction energy between sub-units, relative to the interaction strength, $\left<U_{ss}/\epsilon_{ss} \right>$. (c) The average total interaction energy between sub-units and the membrane, relative to the interaction strength, $\left<U_{ms}/\epsilon_{ms}\right>$.}
\end{center}
\end{figure*}

First, in Fig.~\ref{fig:yield_time_visc_0.1_kap_2}, we plot the averages of various quantities as a function of $\epsilon_{ms}$ and $\epsilon_{ss}$ at different simulation times, $t$, for $\lambda_b = 2\sqrt{3} k_BT$ and $\eta_m = 133.3 m/t_0$. Considering Fig.~\ref{fig:yield_time_visc_0.1_kap_2}(a), at later times, the largest number of correctly assembled cores are obtained for the second highest $\epsilon_{ss}$, $7.38k_BT$. This is close to the optimal value obtained in previous work~\cite{wilber} with a very similar model of about $7.14k_BT$. Although for the highest value, $\epsilon_{ss} = 8.02k_BT$, the total interaction energy between sub-units relative to the interaction strength is somewhat lower, see Fig.~\ref{fig:yield_time_visc_0.1_kap_2}(b), this corresponds to many incomplete cores assembling, thus starving the system of free sub-units. This kinetic trap is not related to the membrane and has often been observed previously~\cite{hagan2013}. In contrast, for high $\epsilon_{ss}$, increasing attraction to the membrane hinders complete assembly due to the membrane enveloping, or partly surrounding, partial cores too quickly, thus preventing sub-units or other partial cores from approaching them. The fast envelopment is apparent in Fig.~\ref{fig:yield_time_visc_0.1_kap_2}, where it may be seen that, for high $\epsilon_{ms}$ and $\epsilon_{ss}$, the interaction energy of the sub-units with the membrane approaches its final value much more quickly than the yield. Similarly to the MC results, a promotion of the finite-time assembly for low $\epsilon_{ss}$ at high $\epsilon_{ms}$ occurs. 

Comparing results for equilibrium, Fig.~\ref{fig:MC_yield}, to those from dynamical simulations, there are several differences. Clearly, for equilibrium results, kinetic traps do not play a role. Furthermore, the interaction strength at which assembly starts is slightly lower at equilibrium than after a finite time in dynamical simulation. For the lowest $\epsilon_{ms}$, $\epsilon_{ms} = 0.12k_BT$, where the membrane does not play a significant role and assembly occurs in the bulk, whereas at equilibrium there are complete cores at $\epsilon_{ss} = 5.46 k_BT$, albeit at a relatively small yield, no complete cores were formed within the allowed time in dynamical simulations. Similarly, the range of parameters for which there is assembly promotion is larger for the equilibrium results.

\begin{figure}[ht]
\begin{center}

\includegraphics[scale=0.4]{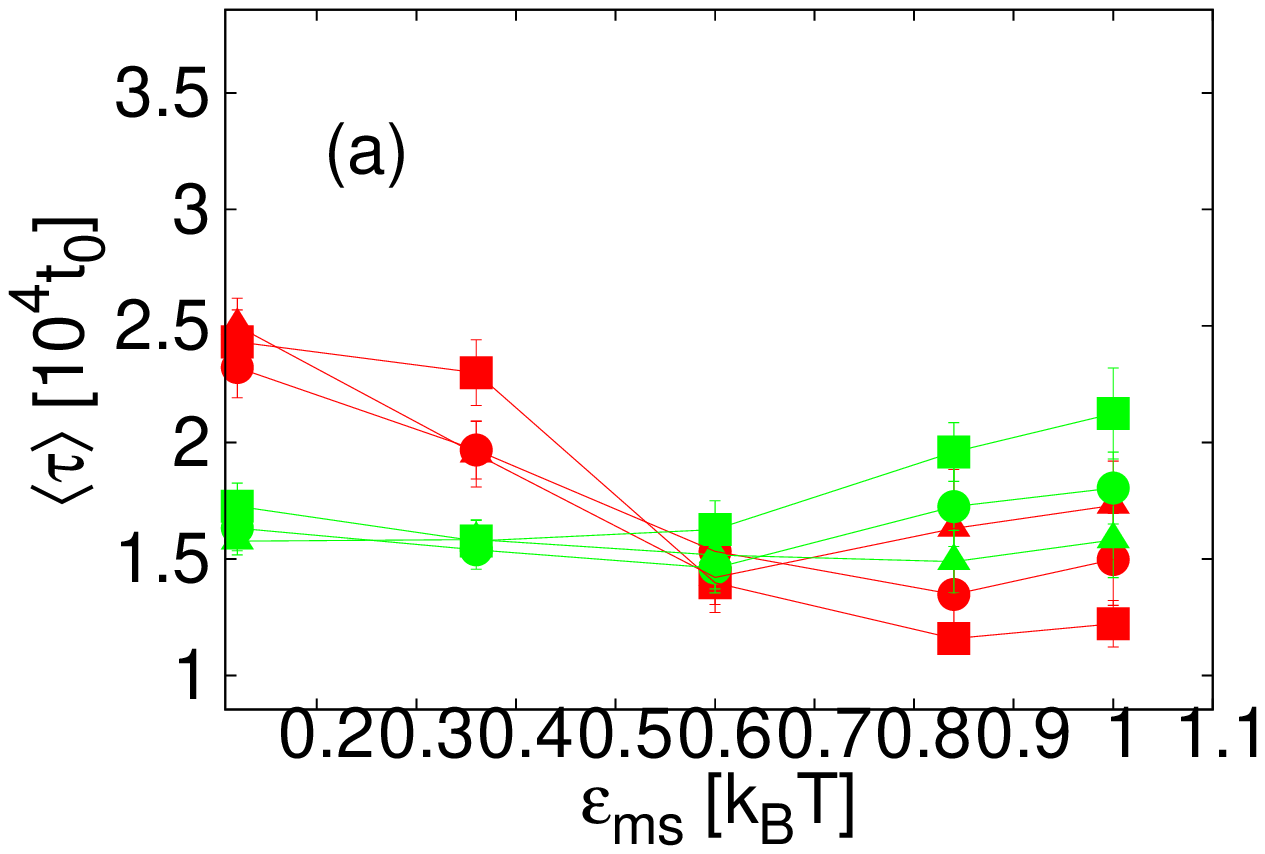}
\includegraphics[scale=0.4]{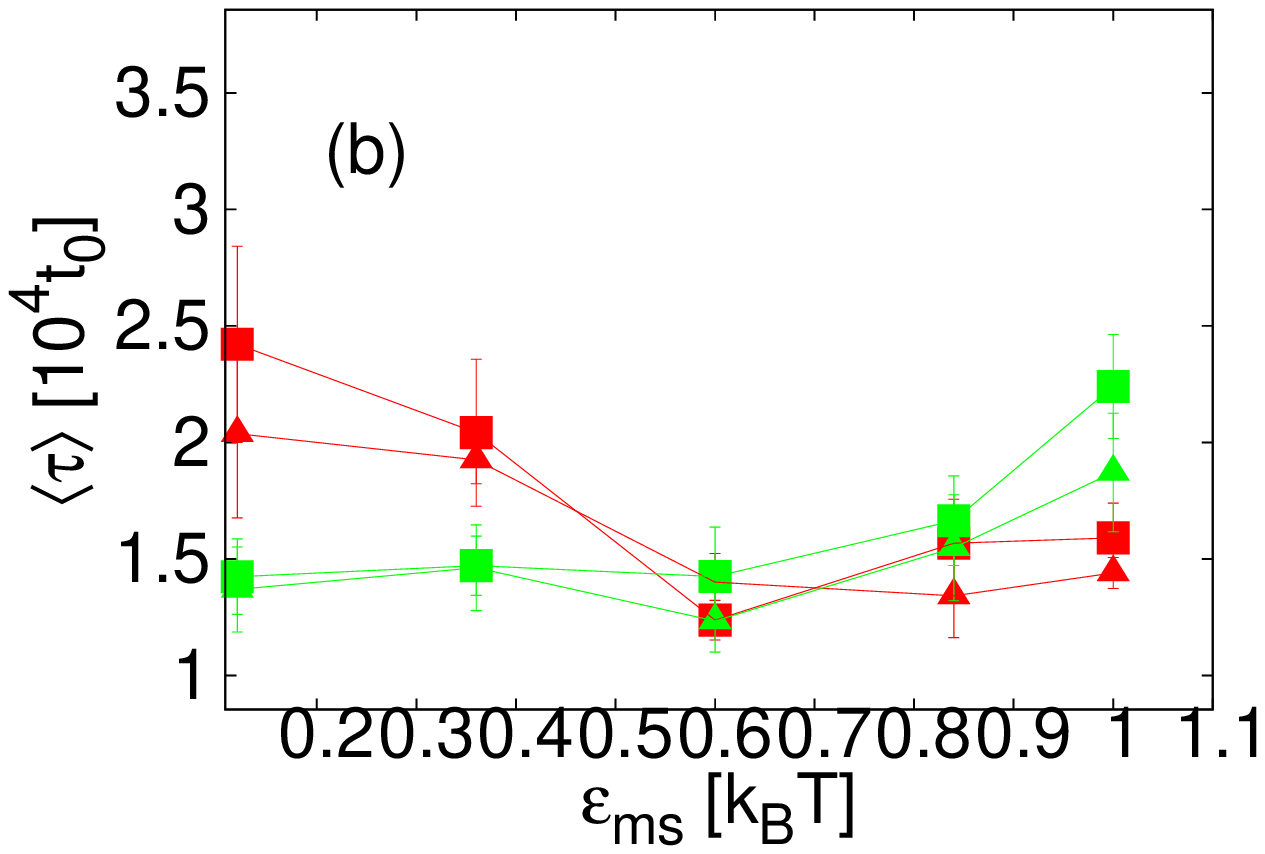}

\caption{\label{fig:first_core_time} Average time until the first complete core is assembled, $\left<\tau\right>$, as a function of membrane-sub-unit interaction strength, $\epsilon_{ms}$, for inter-sub-unit interaction strength, $\epsilon_{ss} = 6.42k_BT$ (red) and $\epsilon_{ss} = 7.38k_BT$ (green), with different membrane viscosities: $\eta_m = 35.1 m/t_0$ ($\blacktriangle$); $\eta_m = 133.3 m/t_0$ ($\CIRCLE$); $\eta_m = 1190 m/t_0$ ($\blacksquare$). (a) Membrane stiffness, $\lambda_b = 2\sqrt{3}k_BT$. (b) $\lambda_b = 4\sqrt{3} k_BT$.}
\end{center}
\end{figure}

Considering  Fig.~\ref{fig:yield_time_visc_0.1_kap_2}(a), we note that a time lag before complete cores are assembled, seen in previous work~\cite{hagan2013}, is apparent for many data points at $t = 10^4 t_0$, including for $\epsilon_{ss} = 7.38 k_BT$ and low $\epsilon_{ms}$, where the yield is highest at later times. However, for some data points, primarily with high $\epsilon_{ms}$, some complete cores are already present at $t = 10^4 t_0$: as well as causing a higher yield for low $\epsilon_{ss}$ once assembly has progressed significantly, attraction to the membrane may also speed up the formation of a single core. By confining sub-units to a surface, the effective size of the space that they must search to find each other is reduced. The membrane may also mediate effective attractions, directing sub-units and partial cores towards each other~\cite{reynwar2007} and, if deformation occurs, it may bring membrane-attached sub-units closer together. Conversely, deformation of the membrane may also tend to block assembly, preventing partial cores from being accessed by sub-units or other partial cores, leading to an increase in assembly time. The effect of the membrane on single-core assembly times is also shown in Fig.~\ref{fig:first_core_time}, where we plot the average time until the first complete core in the system is formed, which we denote $\left<\tau\right>$, as a function of $\epsilon_{ms}$. We note that, since this quantity is based on a single assembly event, large fluctuations were seen for lower interaction strengths and for some parameters additional simulations were run. For both membrane stiffnesses shown, $\lambda_b = 2\sqrt{3}k_BT$ and $4\sqrt{3}k_BT$, $\left<\tau\right>$ tends to be lower for high $\epsilon_{ms}$ for $\epsilon_{ss} = 6.42 k_BT$, whereas for $\epsilon_{ss} = 7.38 k_BT$ the curve is flatter. For sub-unit interaction strengths that are approximately optimal for bulk assembly, the process is sufficiently fast that the membrane does not affect $\left<\tau\right>$, whereas for lower values it may cause a significant speed-up. However, we note that, for high $\epsilon_{ms}$ with some parameters, $\left<\tau\right>$ shows an increase. This is consistent with the membrane blocking assembly.

\begin{figure}[ht]
\begin{center}

\includegraphics[scale=0.32]{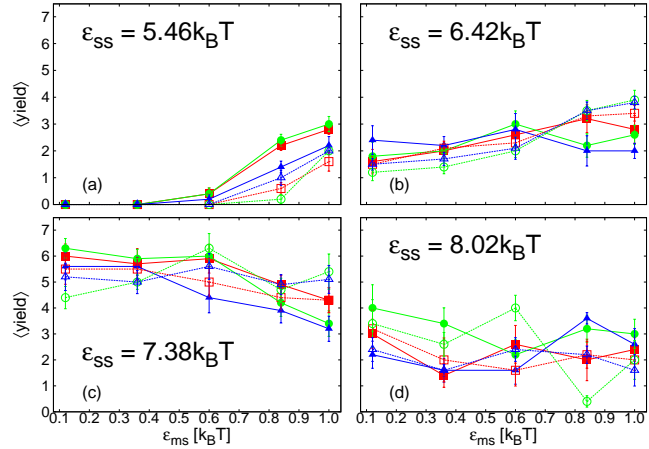}

\caption{\label{fig:yield_eps_ms_kap_1} Plots of the average yield of complete cores, $\left<yield\right>$, as a function of membrane-sub-unit interaction strength, $\epsilon_{ms}$, for membrane stiffness, $\lambda_b = \sqrt{3} k_BT$ at $t = 5 \times 10^4 t_0$. From simulations with SRD (solid lines, filled symbols) or LD (dashed lines, open symbols), for different membrane viscosities: $\eta_m = 35.1 m/t_0$ (blue, $\blacktriangle$/$\vartriangle$); $\eta_m = 133.3 m/t_0$ (green, $\CIRCLE$/$\Circle$); $\eta_m = 1190 m/t_0$ (red, $\blacksquare$/$\square$); and sub-unit interaction strengths, $\epsilon_{ss}$, as indicated on the panels (a) - (d).}
\end{center}
\end{figure}

\begin{figure}[ht]
\begin{center}

\includegraphics[scale=0.32]{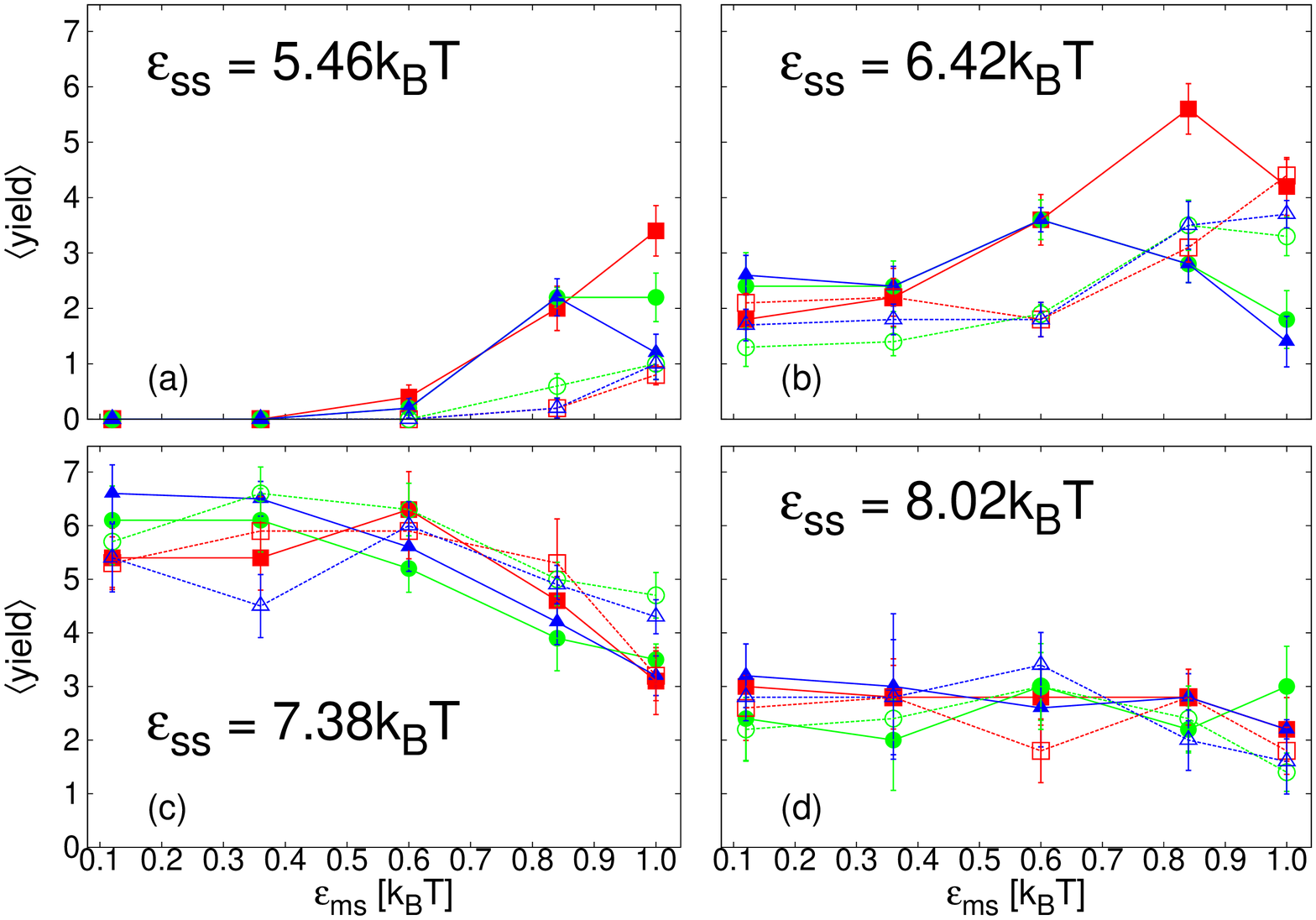}

\caption{\label{fig:yield_eps_ms_kap_2} Plots of the average yield of complete cores, $\left<yield\right>$, as a function of membrane-sub-unit interaction strength, $\epsilon_{ms}$, for membrane stiffness, $\lambda_b =  2\sqrt{3} k_BT$ at $t = 5 \times 10^4 t_0$. From simulations with SRD (solid lines, filled symbols) or LD (dashed lines, open symbols), for different membrane viscosities: $\eta_m = 35.1 m/t_0$ (blue, $\blacktriangle$/$\vartriangle$); $\eta_m = 133.3 m/t_0$ (green, $\CIRCLE$/$\Circle$); $\eta_m = 1190 m/t_0$ (red, $\blacksquare$/$\square$); and sub-unit interaction strengths, $\epsilon_{ss}$, as indicated on the panels (a) - (d).}
\end{center}
\end{figure}

\begin{figure}[ht]
\begin{center}

\includegraphics[scale=0.32]{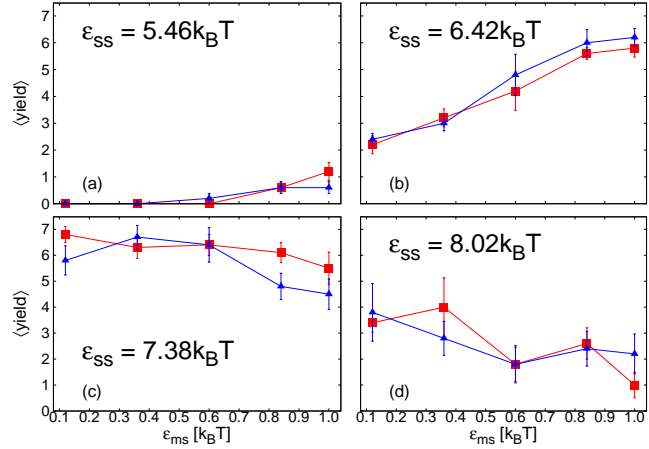}

\caption{\label{fig:yield_eps_ms_kap_4} Plots of the average yield of complete cores, $\left<yield\right>$, as a function of membrane-sub-unit interaction strength, $\epsilon_{ms}$, for membrane stiffness, $\lambda_b =  4\sqrt{3} k_BT$ at $t = 5 \times 10^4 t_0$. From simulations with SRD, for different membrane viscosities: $\eta_m = 35.1 m/t_0$ (blue, $\blacktriangle$/$\vartriangle$); $\eta_m = 1190 m/t_0$ (red, $\blacksquare$/$\square$); and sub-unit interaction strengths, $\epsilon_{ss}$, as indicated on the panels (a) - (d).}
\end{center}
\end{figure}

We next consider the average yield of complete cores, $\left<yield\right>$, measured at the end of the simulation, at time $t = 5 \times 10^4 t_0$. We present results for all the different parameter sets we have simulated, except for the lowest $\epsilon_{ss}$, for which only a very small amount of assembly occurred at the highest $\epsilon_{ms}$. We split the results into 3 figures by membrane stiffness: $\lambda_b = \sqrt{3}k_BT$ in Fig.~\ref{fig:yield_eps_ms_kap_1}; $\lambda_b = 2\sqrt{3}k_BT$ in Fig.~\ref{fig:yield_eps_ms_kap_2}; $\lambda_b = 4\sqrt{3}k_BT$ in Fig.~\ref{fig:yield_eps_ms_kap_4}. Within each figure, results are divided by $\epsilon_{ss}$ into four different sub-figures labelled (a) - (d). $\left<yield\right>$ is plotted as a function of $\epsilon_{ms}$, with curves corresponding to different membrane viscosities and simulation methods indicated by different colors, symbols and line types. Rather than describe in detail the specific features of each figure, we discuss the general trends that arise and pick out interesting features.

Overall, as $\epsilon_{ss}$ is increased, we identify three different trends for finite-time assembly with realistic dynamics. Firstly, at $\epsilon_{ss} = 5.46k_BT$ and $6.42 k_BT$, both lower than the bulk-assembly optimal value, increasing $\epsilon_{ms}$, which tends to increase the membrane deformation rate in all regimes, may promote finite-time assembly. With high $\epsilon_{ms}$, assembly also occurs for values of $\epsilon_{ss}$ where there is no bulk assembly within our simulation time. However, the rate at which the membrane deforms is important. It is influenced by various factors, for example the strength of the attraction of sub-units to the membrane or membrane viscosity, and results in competing effects on assembly. Increasing it via $\epsilon_{ms}$ may at first aid assembly, as seen in the initial increase in $\left<yield\right>$ with $\epsilon_{ms}$ in parts (a) and (b) of Figs.~\ref{fig:yield_eps_ms_kap_1} -~\ref{fig:yield_eps_ms_kap_4}, but when it is too high the yield may decrease again, as is seen particularly clearly in Fig.~\ref{fig:yield_eps_ms_kap_2}(a) and (b). Results depend on factors such as membrane viscosity and hydrodynamic interactions: decreasing membrane viscosity or including hydrodynamic interactions both increase the deformation rate. Attraction to, deformation of, and encapsulation within, the membrane is expected to decrease the sub-unit entropy such that the difference in entropy between unassembled and assembled states is less. Additionally, the deformation of the membrane may help to guide sub-units attached to it towards each other. It is expected that these effects will all play a role in promoting assembly, although their relative importance may not be easily deduced. However, if the deformation occurs too quickly, before complete cores are formed, the membrane will hinder further sub-units, or other partial cores, from approaching the partial structure, preventing its completion. 

Interestingly, in this low $\epsilon_{ss}$ regime, finite-time assembly is promoted even for $\lambda_b = 4\sqrt{3} k_BT$, Fig.~\ref{fig:yield_eps_ms_kap_4}(a)-(b), although for this membrane stiffness budding only occurs for the highest $\epsilon_{ms}$. For this membrane stiffness, results do not depend on membrane viscosity, confirming that here budding does not play a role. Despite a lack of  envelopment, it is expected that attachment to the membrane will nonetheless reduce the entropic cost of forming a partial core. A second plausible mechanism is a local increase in sub-unit density near the membrane surface. In contrast, for the lower two $\lambda_b$, Fig.~\ref{fig:yield_eps_ms_kap_1}(a)-(b) and Fig.~\ref{fig:yield_eps_ms_kap_2}(a)-(b), the membrane deformation rate does play a role. Comparing SRD and LD results, simulations with hydrodynamic interactions show a larger promotion of finite-time assembly as $\epsilon_{ms}$ is increased, at least initially. For some parameters the yield decreases again as $\epsilon_{ms}$ is increased further and this drop off occurs earlier with hydrodynamic interactions. Furthermore, particularly for $\lambda_b = 2\sqrt{3} k_BT$, membrane viscosity, $\eta_m$, is also important. Especially for SRD results, decreasing $\eta_m$ shifts the point at which the finite-time yield begins to decrease to lower $\epsilon_{ms}$. Interestingly, the effect of hydrodynamic interactions and membrane viscosity are much stronger for $\lambda_b = 2\sqrt{3} k_BT$ than for $\lambda_b = \sqrt{3} k_BT$.

In the second regime, for $\epsilon_{ss} = 7.38k_BT$, at about the bulk-assembly optimal value, increasing $\epsilon_{ms}$ tends to decrease the finite-time yield. Here, there is no strong dependence on membrane viscosity or hydrodynamic interactions and, additionally, results are quite similar for all three membrane stiffnesses. This suggests membrane deformation is not crucial, rather the decrease in yield may occur because attraction to the membrane promotes the faster assembly of partial cores, bringing the system into the monomer starvation trap that is only seen in the bulk for higher $\epsilon_{ss}$.

Finally, for the highest $\epsilon_{ss}$, $\epsilon_{ss} = 8.02k_BT$, where there is a monomer starvation kinetic trap for bulk assembly, there is no clear effect of the membrane on finite-time assembly. Since assembly, at least of partial cores, occurs very quickly in the bulk, results here are likely dominated by non-membrane-associated assembly.

\begin{figure}[ht]
\begin{center}

\includegraphics[scale=0.32]{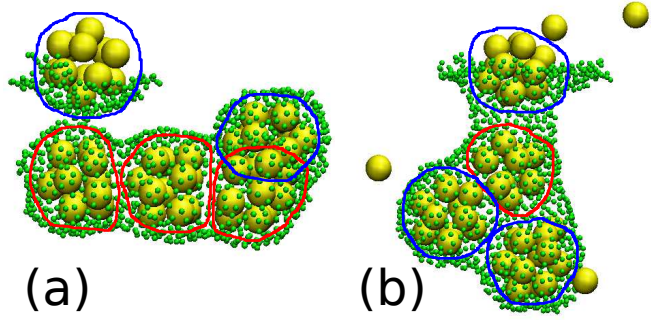}

\caption{\label{fig:dynam_assemb_mem_snapshots} Snapshots from simulations with SRD at $t = 5 \times 10^4 t_0$ with membrane stiffness, $\lambda_b = 2\sqrt{3} k_BT$ and membrane viscosity, $\eta_m = 133.3 m/t_0$: (a) Membrane-sub-unit interaction strength, $\epsilon_{ms} = k_BT$ and inter-sub-unit interaction strength, $\epsilon_{ss} = 7.38 k_BT$; (b) $\epsilon_{ms} = 0.6 k_BT$, $\epsilon_{ss} = 6.42 k_BT$. Sub-units are shown in yellow and membrane particles in green. Only sub-units within $6 l_0$ of a membrane particle are plotted. Membrane particle size has been reduced to make structures within buds more visible. Completed cores are circled in blue, whilst partially assembled ones are circled in red.}
\end{center}
\end{figure}

In Fig.~\ref{fig:dynam_assemb_mem_snapshots} we show snapshots of final configurations from simulations with $\lambda_b = 2\sqrt{3} k_BT$ and $\eta_m = 133.3 m/t_0$. For $\epsilon_{ms} = k_BT$ and $\epsilon_{ss} = 7.38 k_BT$, Fig.~\ref{fig:dynam_assemb_mem_snapshots}(a), for which the average yield was reduced compared to the low $\epsilon_{ms}$ value, three of the four cores encapsulated in a bud that are depicted are incomplete. This snapshot corresponds to the optimal $\epsilon_{ss}$ regime. In contrast, for $\epsilon_{ms} = 0.6 k_BT$ and $\epsilon_{ss} = 6.42 k_BT$, Fig.~\ref{fig:dynam_assemb_mem_snapshots}(b), for which the average yield was enhanced compared to low $\epsilon_{ms}$ value, only one of the four cores encapsulated, or partially encapsulated, in a bud that are depicted is incomplete. This snapshot corresponds to the low $\epsilon_{ss}$ regime. Although these snapshots only depict the situation in two individual runs, they illustrate how the membrane may block assembly completion when its deformation rate is too high by preventing partial cores from being accessed by sub-units or other partial cores.

\begin{figure}[ht]
\begin{center}
\begin{tabular}{cc}
\includegraphics[scale=0.33]{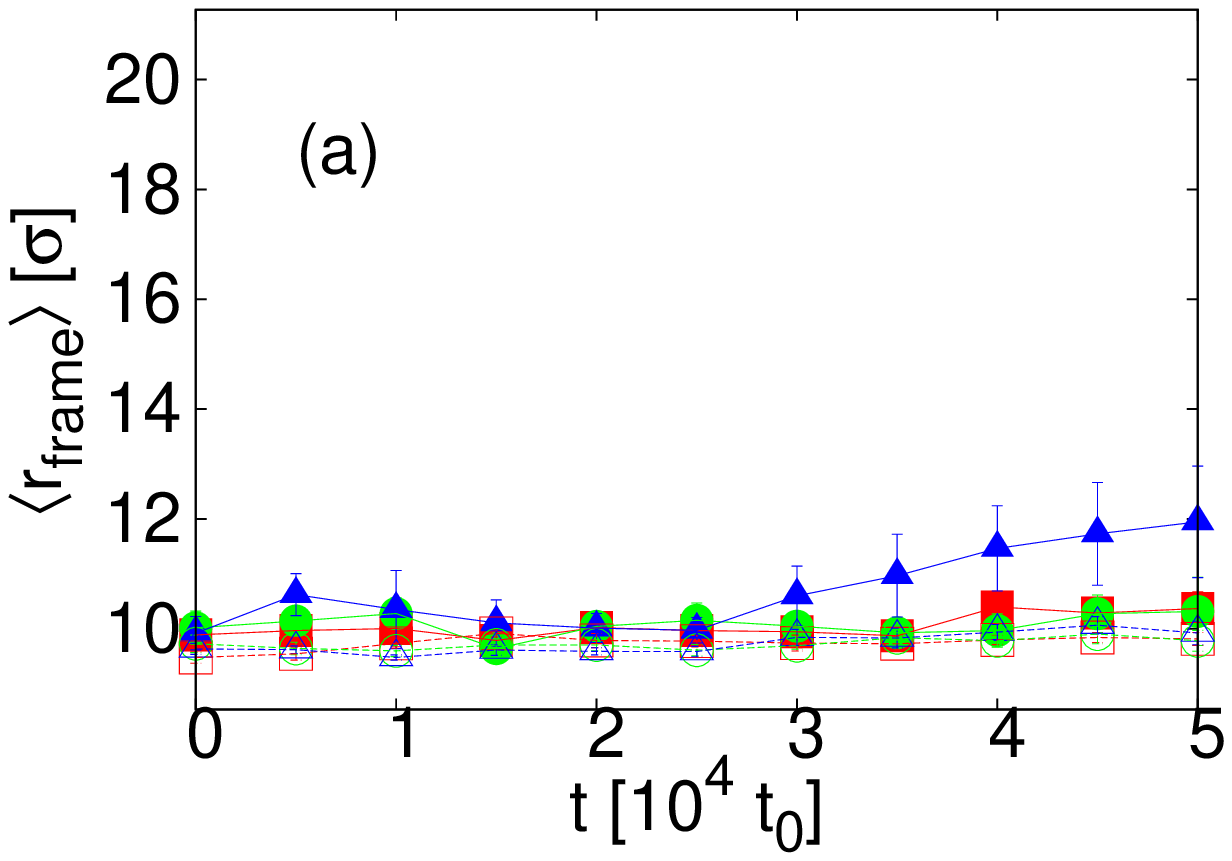} &
\includegraphics[scale=0.33]{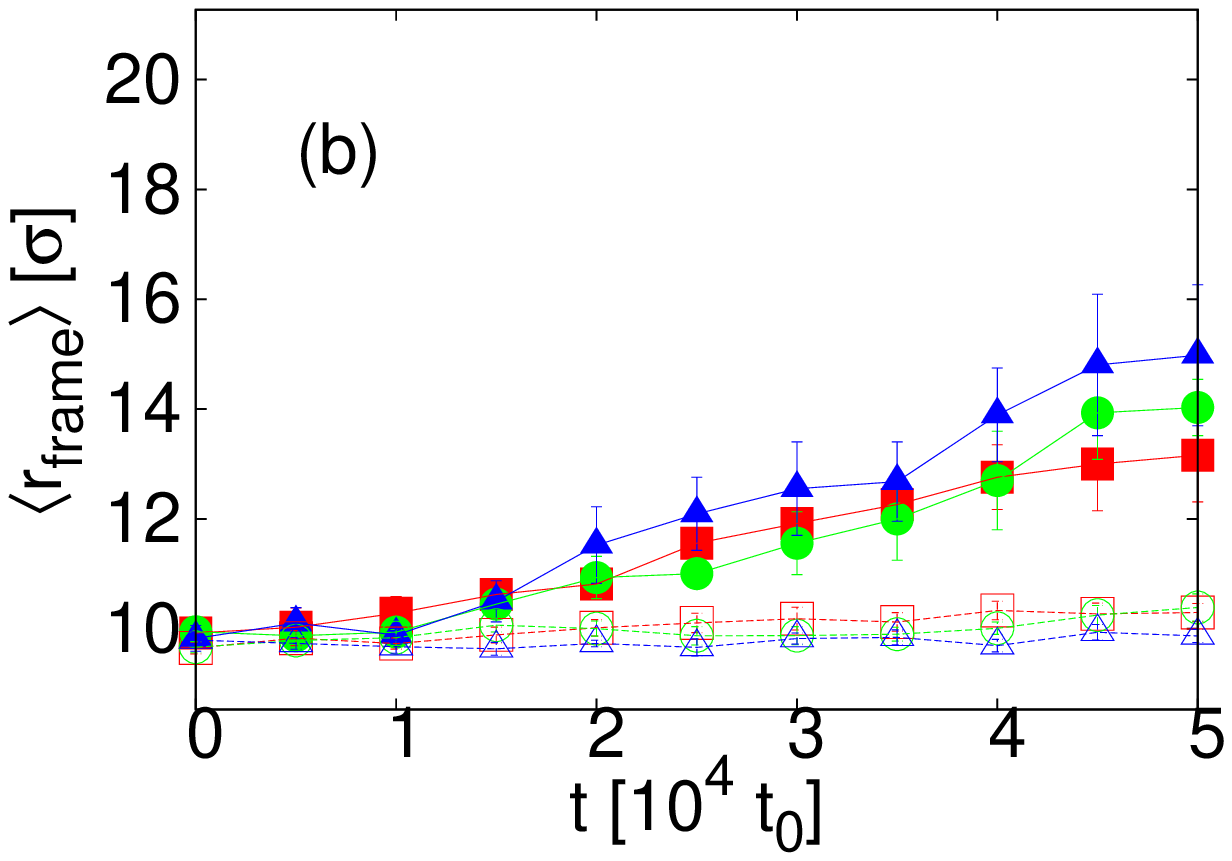} \\
\includegraphics[scale=0.33]{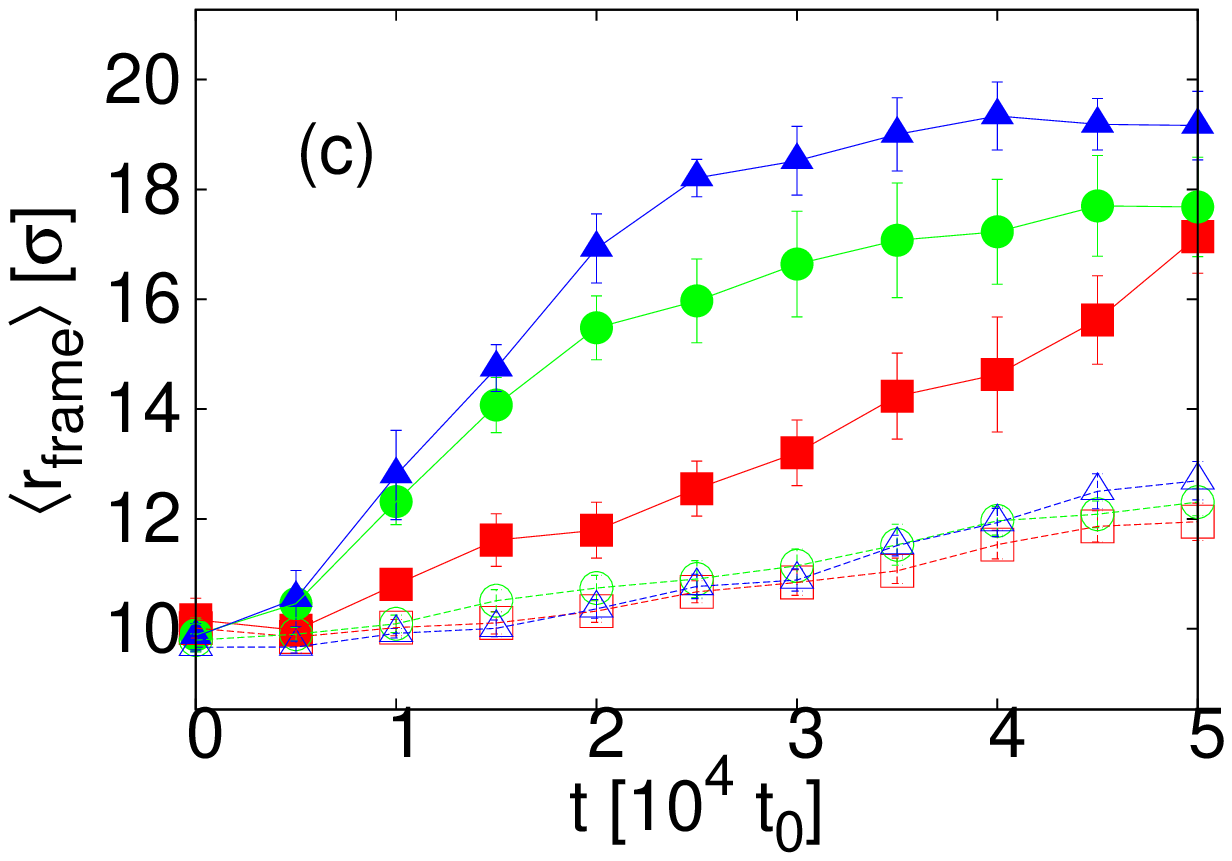} &
\includegraphics[scale=0.33]{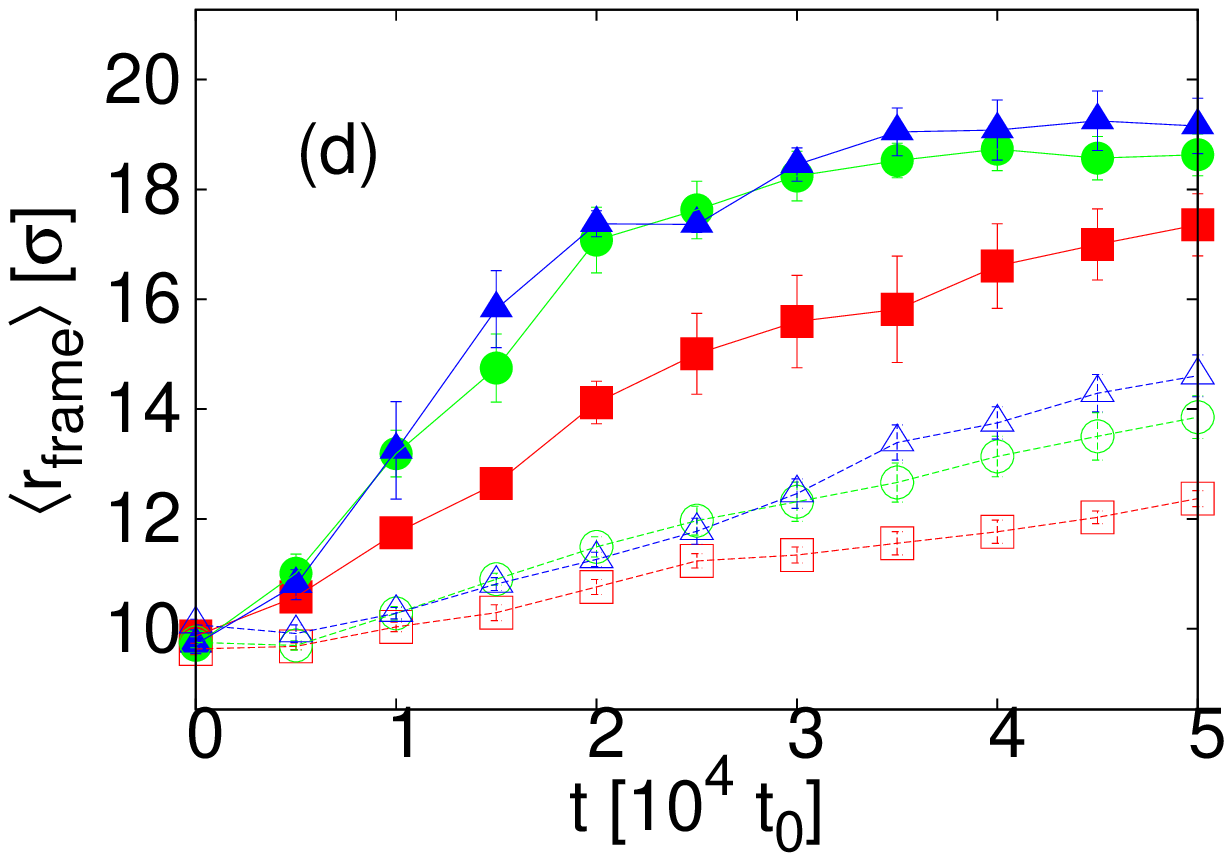} \\
\end{tabular}
\caption{\label{fig:r_frame_time} Plots of the average position of the frame, $\left<r_{frame}\right>$,  as a function of time, $t$, for membrane stiffness, $\lambda_b = 2\sqrt{3} k_BT$ and inter-sub-unit interaction strength, $\epsilon_{ss} = 6.42 k_BT$. From simulations with SRD (solid lines, filled symbols) or LD (dashed lines, open symbols), for different membrane viscosities: $\eta_m = 35.1 m/t_0$ (blue, $\blacktriangle$/$\vartriangle$); $\eta_m = 133.3 m/t_0$ (green, $\CIRCLE$/$\Circle$); $\eta_m = 1190 m/t_0$ (red, $\blacksquare$/$\square$); and sub-unit-membrane interaction strengths: (a) $\epsilon_{ms} = 0.36 k_BT$; (b) $\epsilon_{ms} = 0.6 k_BT$; (c) $\epsilon_{ms} = 0.84 k_BT$; (d) $\epsilon_{ms} = k_BT$.}
\end{center}
\end{figure}

A useful quantity to indicate the amount of membrane deformation is $r_{frame}$, the distance from the edge of the simulation box of the frame to which the edge of the membrane is bound, which increases as the membrane distorts its shape out of the plane. To show how changing membrane viscosity, and including hydrodynamic interactions, alters the rate and extent of membrane deformation, we plot, in Fig.~\ref{fig:r_frame_time}, $\left<r_{frame}\right>$ as a function of time with $\lambda_b = 2\sqrt{3}k_BT$ and $\epsilon_{ss} = 6.42 k_BT$ for different $\epsilon_{ms}$ from $0.36k_BT$. Apart from the lowest $\epsilon_{ms}$, the membrane deformation occurs faster and to a greater extent for simulations with hydrodynamics. Hydrodynamic interactions increase the rate of budding. Since budding requires the whole of the membrane to move, correlations mediated by hydrodynamics promote it. Furthermore, for SRD simulations at the highest two $\epsilon_{ms}$, there are also significant differences between membrane viscosities with a trend as expected: $\left<r_{frame}\right>$ is largest for the smallest viscosity.

\begin{figure}[ht]
\begin{center}
\begin{tabular}{cc}
\includegraphics[scale=0.3]{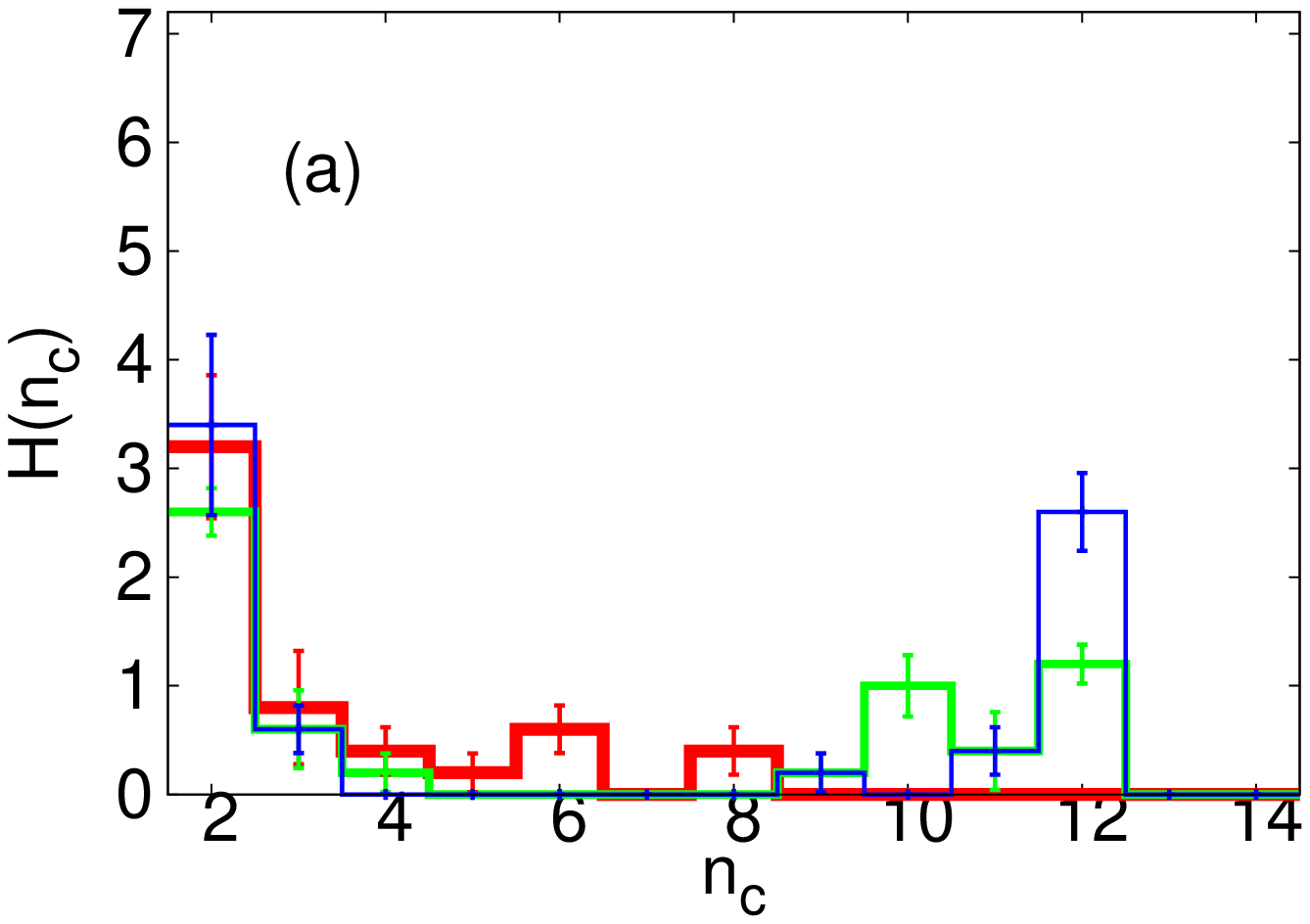}  &
\includegraphics[scale=0.3]{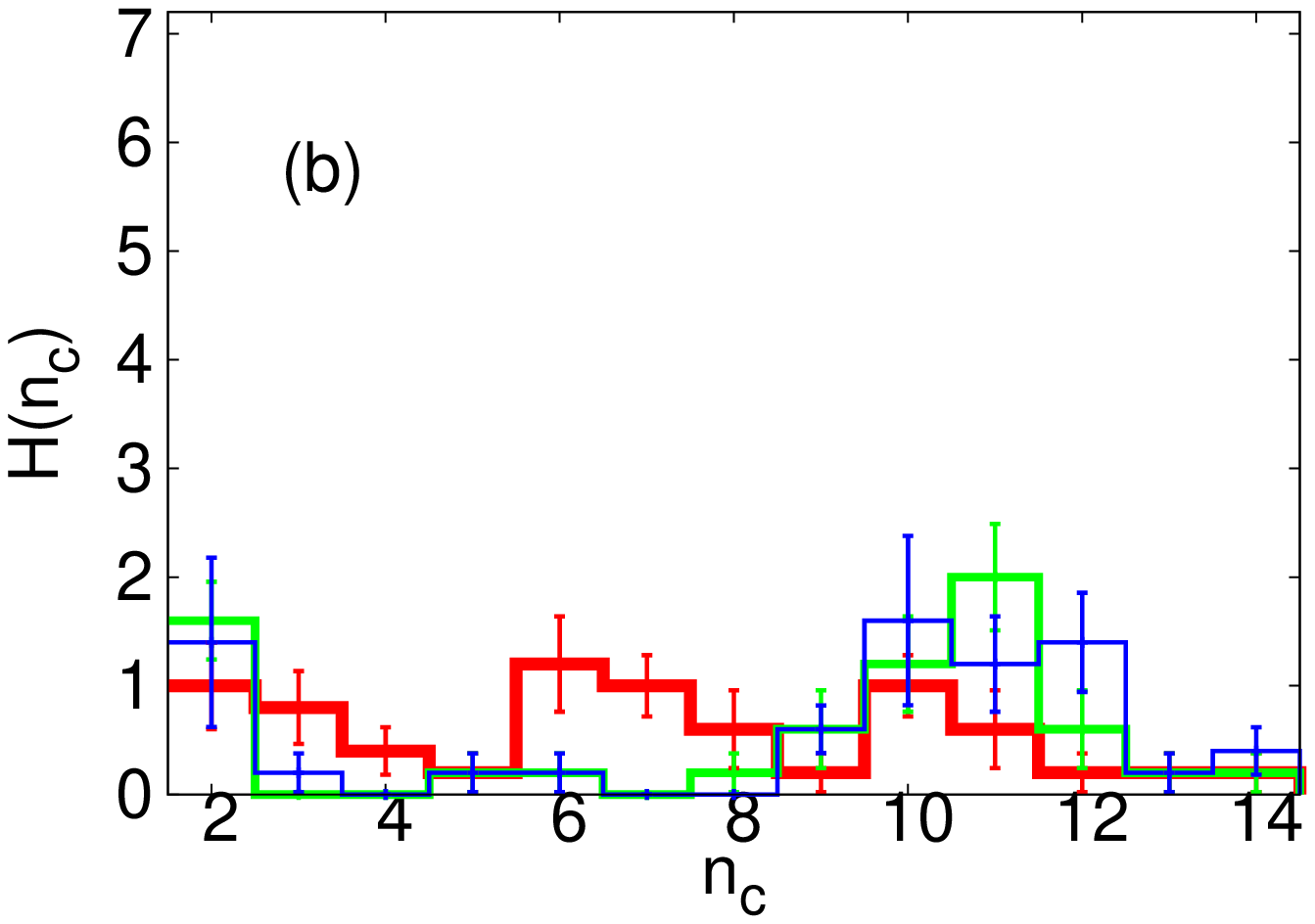}  \\
\includegraphics[scale=0.3]{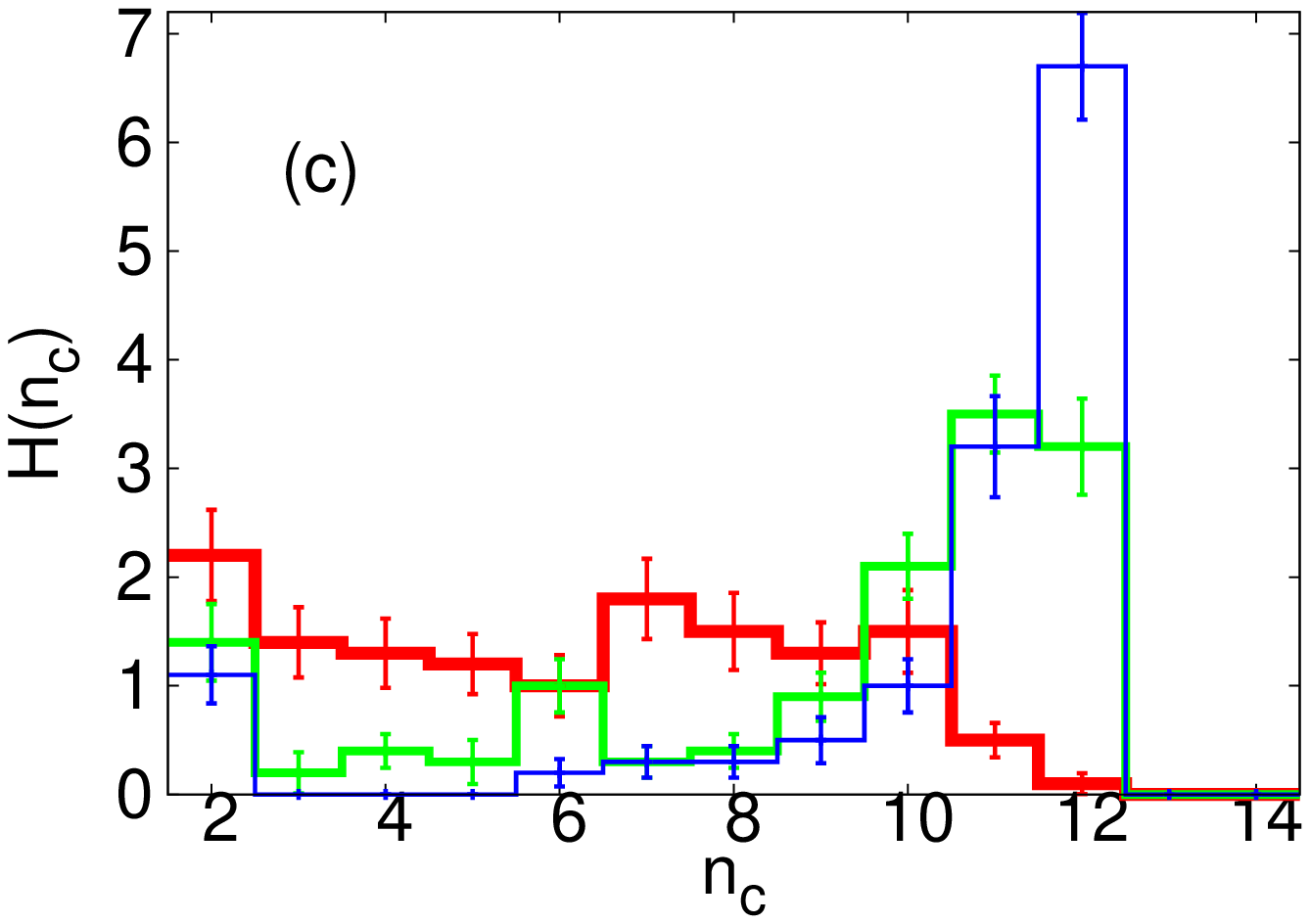}  &
\includegraphics[scale=0.3]{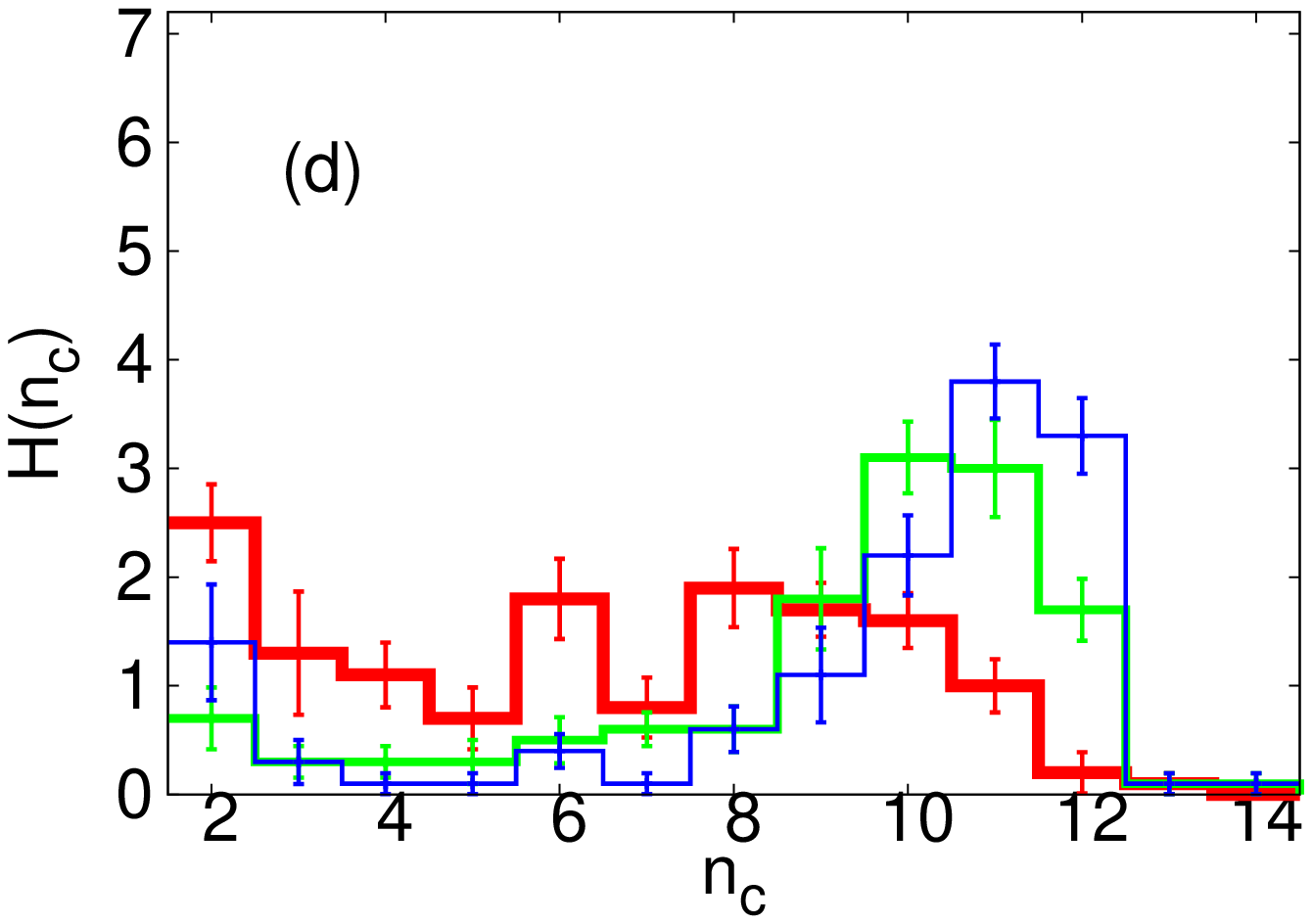}  \\
\end{tabular}
\caption{\label{fig:clust_hists_time} Histograms, $H$, of cluster size, $n_{c}$, for membrane stiffness, $\lambda_b = 2\sqrt{3} k_BT$ and membrane viscosity, $\eta_m = 35.1 m/t_0$ at different times, $t$: $t = 1 \times 10^4 t_0$ (red); $t = 3 \times 10^4 t_0$ (green); $t = 5 \times 10^4 t_0$ (blue); with different parameters: (a) Membrane-sub-unit interaction strength, $\epsilon_{ms} = 0.12 k_BT$ and inter-sub-unit interaction strength, $\epsilon_{ss} = 6.42 k_BT$; (b) $\epsilon_{ms} = k_BT$, $\epsilon_{ss} = 6.42 k_BT$;  (c) $\epsilon_{ms} = 0.12 k_BT$, $\epsilon_{ss} = 7.38 k_BT$; (d) $\epsilon_{ms} = k_BT$, $\epsilon_{ss} = 7.38 k_BT$.}
\end{center}
\end{figure}

We next consider the distributions of cluster size, $n_{c}$, in our simulations, $H \left(n_{c}\right)$. In Fig.~\ref{fig:clust_hists_time}, we plot distributions for $\lambda_b = 2\sqrt{3} k_BT$ and $\eta_m = 35.1 m/t_0$ at three different times. For $\epsilon_{ss} = 6.42k_BT$, the lowest $\epsilon_{ss}$ for which, in dynamical simulations, there is bulk assembly, there are few clusters with intermediate sizes when $\epsilon_{ms}$ is low. At all times considered, the majority of clusters are of size two; at later times there is an additional peak at size twelve, corresponding to complete cores. In contrast, when $\epsilon_{ms}$ is high, the attraction to the membrane stabilizes intermediate cluster sizes at early times. At later times, the clusters have grown but the peak near twelve is less sharp, with similar numbers of cores of size ten and eleven, and also some larger ones. This shows the effect of the membrane blocking the completion of partial cores. It also seen for higher $\epsilon_{ss}$, $\epsilon_{ss} = 7.38k_BT$, where the distribution for low $\epsilon_{ms}$ is much flatter at early times with many clusters of intermediate sizes. 

\begin{figure}[ht]
\begin{center}
\begin{tabular}{cc}
\includegraphics[scale=0.3]{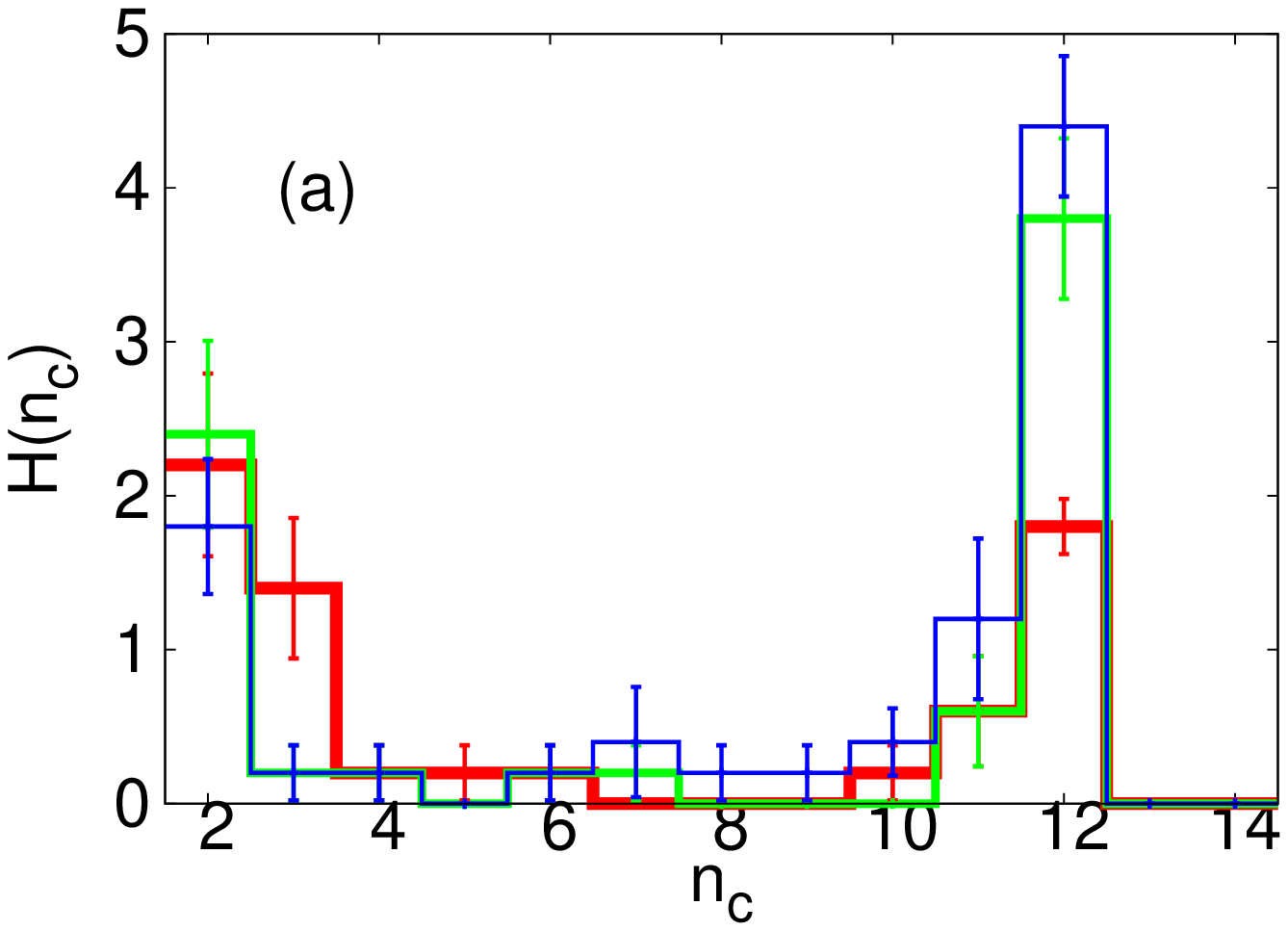}  &
\includegraphics[scale=0.3]{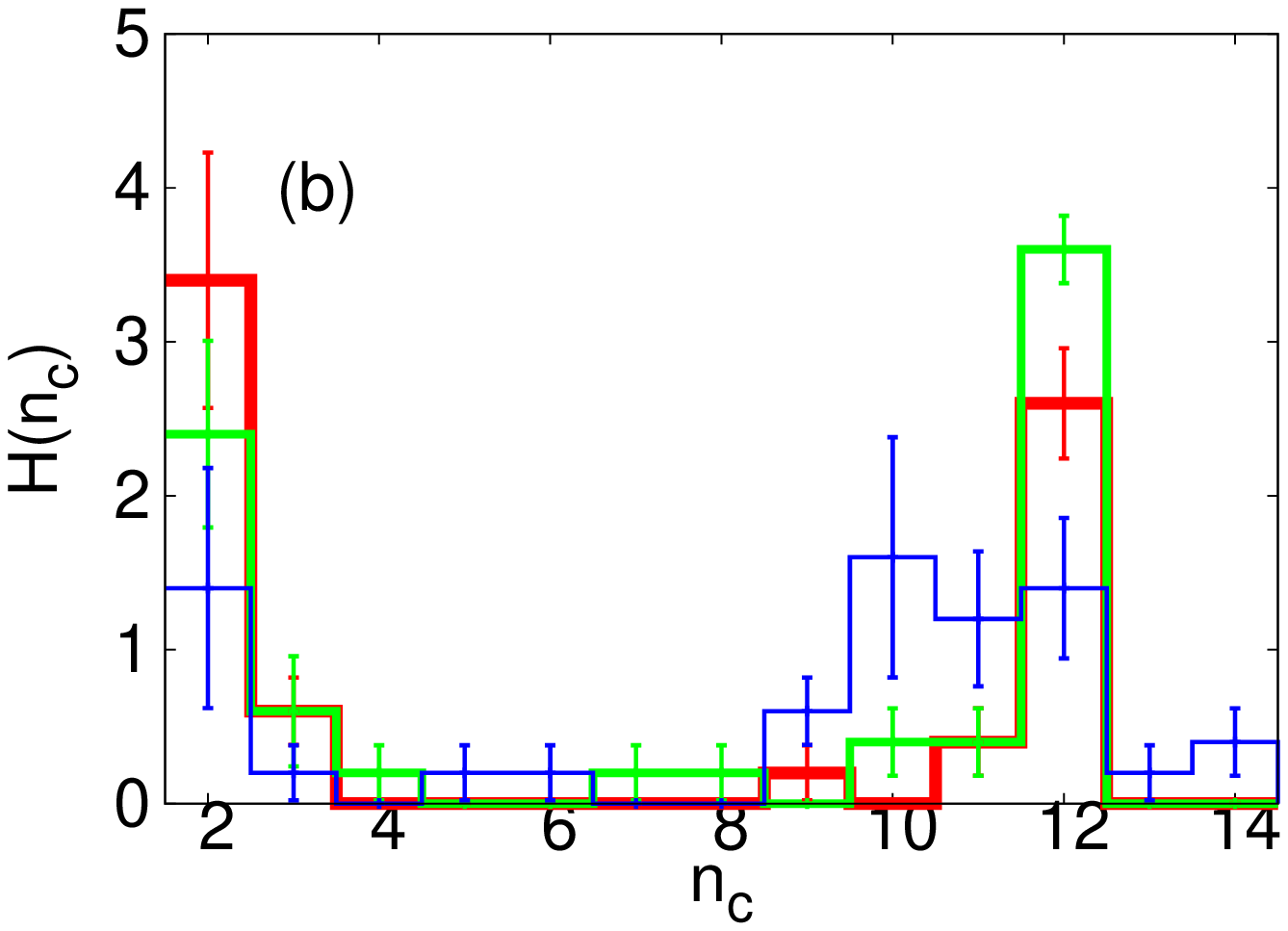}  \\
\includegraphics[scale=0.3]{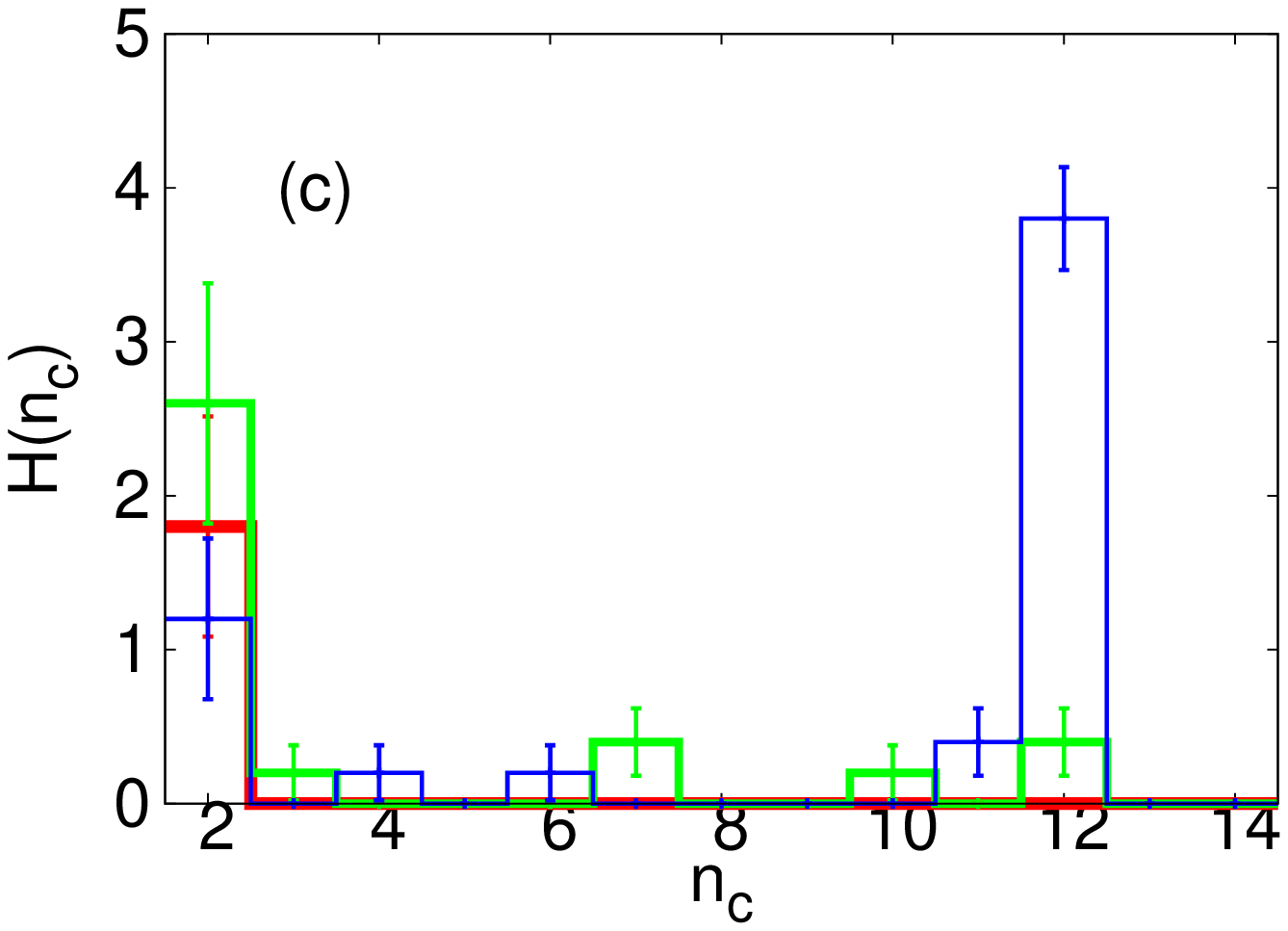}  &
\includegraphics[scale=0.3]{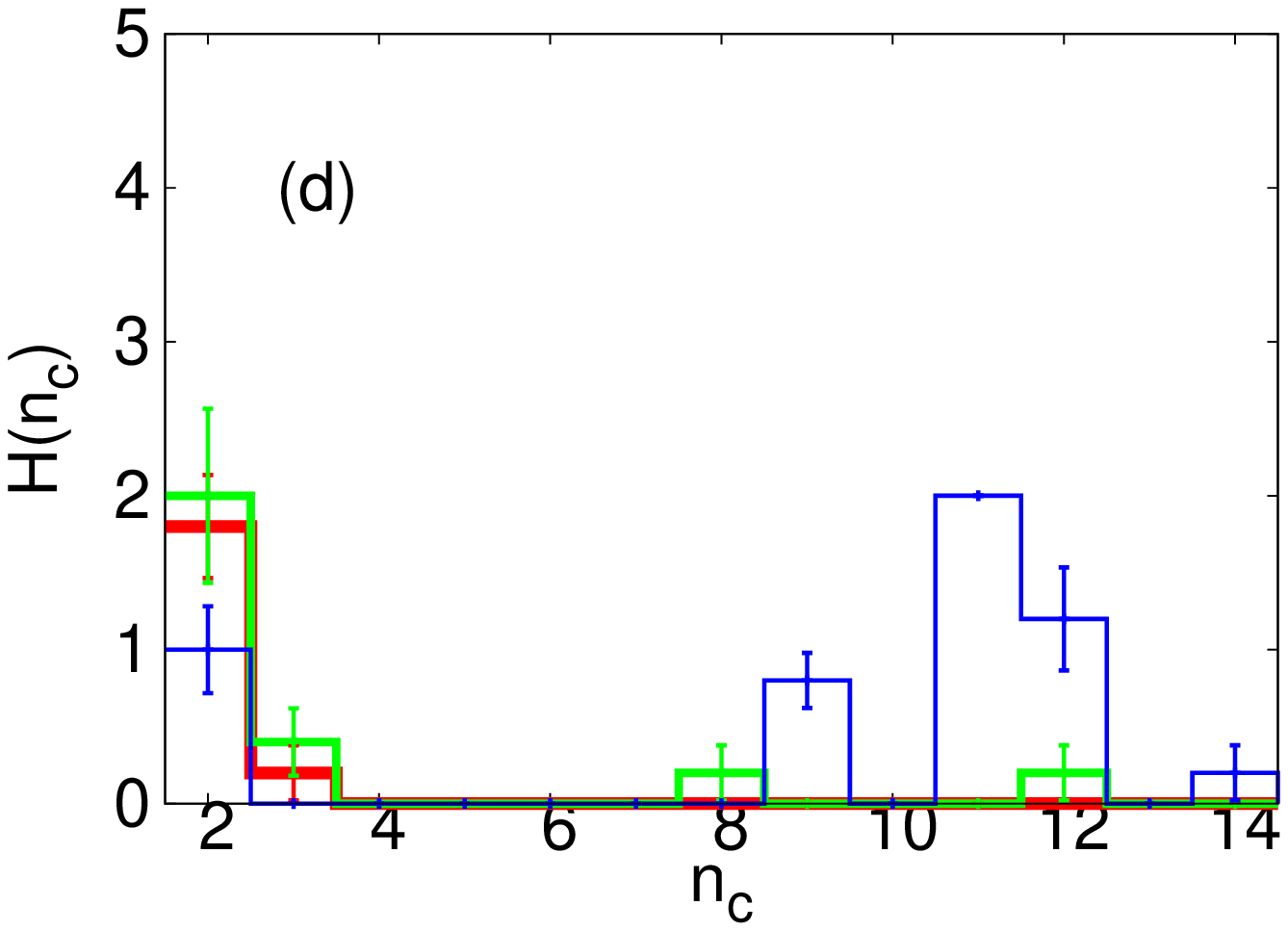}  \\
\end{tabular}
\caption{\label{fig:clust_hists_visc} Histograms, $H$, of cluster size, $n_{c}$, for membrane stiffness, $\lambda_b = 2\sqrt{3} k_BT$ at time, $t = 5 \times 10^4 t_0$ for different membrane-sub-unit interaction strength values: $\epsilon_{ms} = 0.12 k_BT$ (red); $\epsilon_{ms} = 0.6 k_BT$ (green); $\epsilon_{ms} = k_BT$ (blue); with different parameters: (a) Membrane viscosity, $\eta_m = 1190 m/t_0$ and inter-sub-unit interaction strength, $\epsilon_{ss} = 6.42 k_BT$; (b) $\eta_m = 35.1 m/t_0$, $\epsilon_{ss} = 6.42 k_BT$;(c) $\eta_m = 1190 m/t_0$, $\epsilon_{ss} = 5.46 k_BT$; (d) $\eta_m = 35.1 m/t_0$, $\epsilon_{ss} = 5.46 k_BT$.}
\end{center}
\end{figure}

In Fig.~\ref{fig:clust_hists_visc}, we show the effect of membrane viscosity on the cluster size distribution. We plot results from the end of the simulations, at time $t = 5\times 10^4 t_0$, again for $\lambda_b = 2\sqrt{3} k_BT$. With $\epsilon_{ss} = 6.42k_BT$, we see that, for the highest viscosity, increasing $\epsilon_{ms}$ leads to a distribution that is more strongly peaked at twelve. In contrast, for the lowest viscosity, although increasing $\epsilon_{ms}$ does lead to more larger clusters, it also gives a much broader distribution around twelve. A similar effect occurs for $\epsilon_{ss} = 5.46 k_BT$: for this $\epsilon_{ss}$ too, when membrane viscosity is low, high $\epsilon_{ms}$ causes the membrane to encapsulate the assembling capsids too quickly, blocking their completion.

As in similar previous models~\cite{hagan}, the concentration of our sub-units is relatively high compared to experimental systems, and furthermore the number of sub-units in a completed core is low. These choices are necessary for computational tractability but have the consequence that the assembly rates in our simulations are much higher than experimental ones. Assuming that sub-units correspond to capsomers of size on the order of 10nm, and matching the drag coefficient of our sub-units, we estimate that our simulation length is around 5ms, whereas {\it in vitro}~\cite{zlotnick2000} and {\it in vivo}~\cite{baumgartel2012} experiments have observation times on the order of minutes. Thus, a direct quantitative comparison cannot be made. Our results rather demonstrate how the rate of membrane deformation compared to the assembly rate may affect the success of the latter. Furthermore, they show how properties such as membrane viscosity, which might be varied experimentally by changing lipid composition~\cite{espinosa2011} are expected to impact on the assembly process. Since it aids the avoidance of kinetic traps, the interactions between assembly viral capsomers are typically relatively weak~\cite{zlotnick2003}, and thus the first, low $\epsilon_{ss}$ regime identified is mostly likely to be relevant to these systems.

\section{Other target cores}
\label{sec:other_target}

\begin{figure}[ht]
\begin{center}

\includegraphics[scale=0.4]{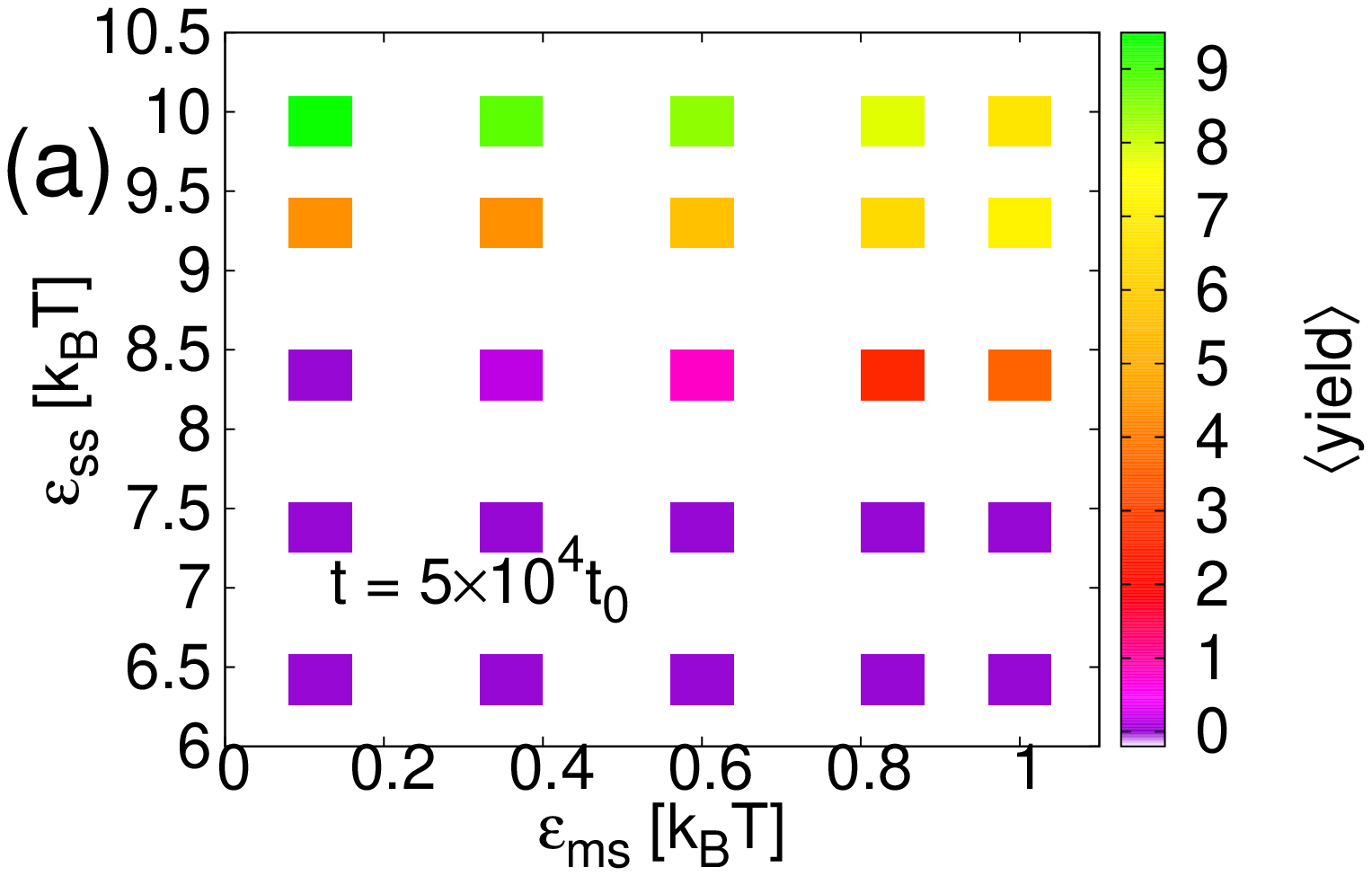}
\includegraphics[scale=0.4]{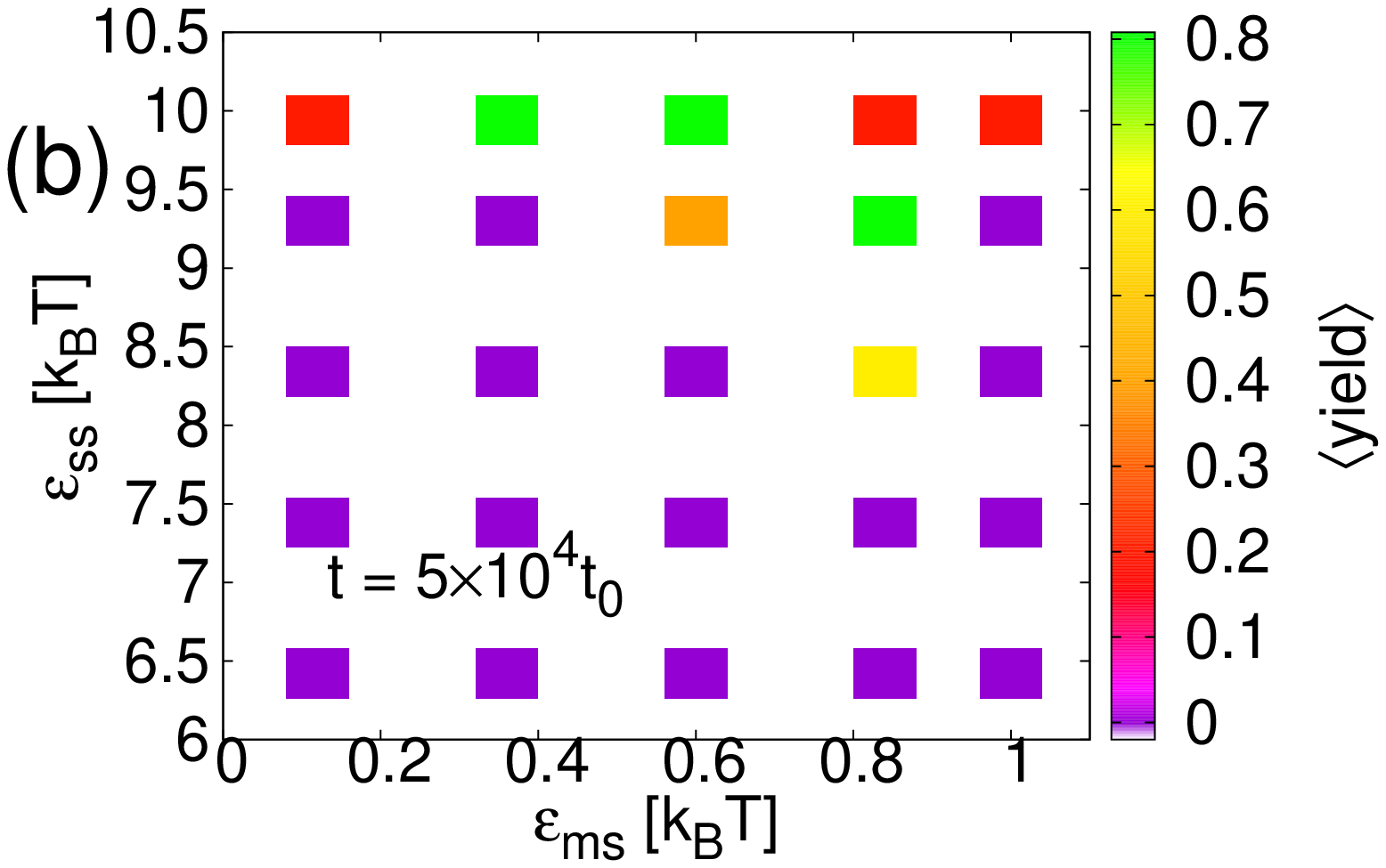}

\caption{\label{fig:yield_OTHER} Plots of the average yield of complete structures, $\left<yield\right>$, as a function of membrane-sub-unit interaction strength, $\epsilon_{ms}$ and inter-sub-unit interaction strength, $\epsilon_{ss}$ at time $t = 5 \times 10^4 t_0$ for membrane stiffness, $\lambda_b = 2\sqrt{3}k_BT$ and membrane viscosity, $\eta_m = 133.3 m/t_0$: (a) Sub-units with interactions to form a cube. (b) Sub-units with interactions to form a dodecahedron. Note the different scales for $\left<yield\right>$ and also the higher values of $\epsilon_{ss}$ as compared to the results for icosahedral cores.}
\end{center}
\end{figure}

We have investigated the effect of a membrane on core assembly of icosahedral cores. It is expected that many of the qualitative features of the results, such as the interplay between the membrane promoting assembly by confining sub-units and hindering it by blocking additional sub-units or other partial cores from approaching partial structures, will be general to other target structures. To gain more insight into the transferability of the findings to other core shapes, we finally, in Fig.~\ref{fig:yield_OTHER}, present results on the yield obtained when the target structure is changed from the icosahedral core. As for icosahedra, we define a cluster to be a complete structure when it contains the correct number of sub-units for the target structure and each sub-unit patch on each sub-unit forms a bond with another member of the cluster. We choose sub-units with patches such that their interactions are minimized for cubic and dodecahedral structures~\cite{wilber}. Otherwise, parameters such as patch width are unchanged, with the membrane patch still lying on the symmetry axis as defined by the sub-unit patches and pointing outwards in a complete structure. In both cases, the sub-units only have three patches for bonding with other sub-units and thus form fewer bonds in a complete structure. Correspondingly, the range of $\epsilon_{ss}$ was shifted up by about $2k_BT$ but the range of $\epsilon_{ms}$ remained the same. 

We simulated for $\lambda_b = 2\sqrt{3} k_BT$ and $\eta_m = 133.3 m/t_0$. For both cubes and dodecahedra, as for icosahedra, budding of the membrane occurred for high $\epsilon_{ms}$. In Fig.~\ref{fig:yield_OTHER}(a), many of the features of the finite-time assembly of icosahedra are reproduced for cubes. For high $\epsilon_{ms}$, there is finite-time assembly at lower $\epsilon_{ss}$ than for low $\epsilon_{ms}$. There is also an increase in finite-time yield with increasing $\epsilon_{ms}$ for the lowest $\epsilon_{ss}$ for which assembly occurs without significant attraction to the membrane. Unlike for icosahedra, at least for these membrane parameters, the yield does not drop off again as $\epsilon_{ms}$ is increased further. This may be because, since they are composed of less sub-units, cubes assemble faster than icosahedra. For the highest $\epsilon_{ss}$, however, there is a reduction in finite-time yield with $\epsilon_{ms}$, similar to results for icosahedra. 

As observed in previous work~\cite{wilber}, finite-time yields of dodecahedra, Fig.~\ref{fig:yield_OTHER}(b), were low. However, here again, there is evidence that attraction to the membrane may promote assembly for $\epsilon_{ss}$ for which it would otherwise not occur. Although it is not apparent in our results, as seen in previous work with a very similar model~\cite{wilber}, it is expected that, if the sub-unit interaction strength were increased sufficiently, the same non-membrane-related kinetic trap that is observed for icosahedra would also be seen for cubes and dodecahedra.

\section{Conclusions}
\label{sec:conc}

To summarize, we have applied a simple patchy-particle model to investigate the effect of interactions with a fluctuating membrane on the dynamics of the assembly of core structures with the same symmetry as many viral cores. As well as interaction strengths, the key parameters we varied were membrane stiffness and viscosity. We also considered the effect of hydrodynamic interactions by simulating both with SRD and LD. As at equilibrium, for assembly with realistic dynamics, attraction to a membrane may promote finite-time assembly, also for sub-unit interaction strengths, $\epsilon_{ss}$, for which it does not occur in the bulk. Furthermore, for $\epsilon_{ss}$ less than the optimal bulk value, attraction to the membrane also decreases the single-core assembly time. 

Membrane budding occurred in dynamically realistic simulations and its rate was strongly increased by hydrodynamic interactions, as well as by lowering the membrane viscosity. The rate of membrane deformation is important in determining the assembly yield after finite time. Relatively high rates may promote assembly by increasing the envelopment of assembling cores and thus decreasing the entropic penalty and also by guiding sub-units towards each other. However, if the rate is too high, the membrane may block partial cores from being completed. Three regimes with different effects of the membrane were identified. For $\epsilon_{ss}$ less than the bulk optimum, finite-time yields depend intricately on a combination of all parameters and may both increase and decrease as attraction to the membrane is increased. For $\epsilon_{ss}$ about equal to the bulk optimum, finite-time yields do not depend strongly on the membrane deformation rate and tend to decrease as attraction to the membrane is increased. For $\epsilon_{ss}$ higher than the bulk optimum, assembly in the bulk is affected by a monomer starvation kinetic trap and the membrane has little influence.

Finally, results with qualitative similarities were also found for core structures with cubic and dodecahedral symmetries. In future work it would be interesting to investigate more different structures, in particular much larger cores.

\begin{acknowledgement}

This work was supported by the Austrian Science Fund (FWF): M1367. Snapshots were created using VMD~\cite{humphrey}. The computational results presented have been achieved in part using the Vienna Scientific Cluster (VSC).

\end{acknowledgement}


\begin{suppinfo}
Definitions of the functions used in inter-particle interactions and the confinement of membrane particles to the frame. In this version the Supporting Information is included below.
\end{suppinfo}

\section{Supporting Information}

We define the various functions used in the interactions between particles and in the confinement of the membrane particles to the frame. The repulsive, $U_{rep}$, and attractive, $U_{att}$, radial potentials used for inter-sub-unit and sub-unit-membrane interactions are given by,
\begin{equation}
U_{rep}(r)= 
\left\{
\begin{array}{l}
4\epsilon\left[\left(\frac{\sigma}{r}\right)^{12}-\left(\frac{\sigma}{r}\right)^{6} + \frac{1}{4} \right]\\
\hspace{1.2 in} \mathrm{for} \; r < r_t,  \\
0\\
\hspace{1.2 in} \mathrm{for} \; r \ge r_t, \\
\end{array}
\right.
\end{equation}
and
\begin{equation}
U_{att}(r)=
\left\{
\begin{array}{l}
-\epsilon\\
\hspace{1.2 in} \mathrm{for} \; r < r_t,\\
4\epsilon\left[\left(\frac{\sigma}{r}\right)^{12}-\left(\frac{\sigma}{r}\right)^{6} \right]\\
\\
\hspace{1.2 in} \mathrm{for} \; r_t  \le r \le r_s,\\
a(r - r_c)^2 + b(r - r_c)^3\\ 
\\
\hspace{1.2 in} \mathrm{for} \; r_s \le r \le r_c,\\
0 \\
\hspace{1.2 in} \mathrm{for} \; r \ge r_c,\\
\end{array}
\right.
\end{equation}
where $r$ is the particle center separation, $r_t = 2^{1/6}\sigma$, $r_s = (\frac{26}{7})^{1/6}\sigma$, $r_c = \frac{67}{48}r_s$, $a = - \frac{24192}{3211}\frac{\epsilon} {r_s^2}$ and $b = -\frac{387072}{61009}\frac{\epsilon}{r_s^3}$. In the range $ r_s \le r \le r_c$, a polynomial interpolation is used for $U_{att}(r)$ so that the potential goes smoothly to 0~\cite{bordat}.

Patchy interactions are produced by multiplying $U_{att}$ by $\gamma_{orient}(\hat{\boldsymbol{r}}_{ij}, \boldsymbol{\Omega}_i,\boldsymbol{\Omega}_j) $, where $\hat{\boldsymbol{r}}_{ij}$ is the unit vector pointing between the particle centers and $ \boldsymbol{\Omega}_i$ an orientation. For inter-sub-unit interactions, $\gamma_{orient}$ is composed of three, and for membrane-sub-unit interactions only one, factor of the following functional form~\cite{miller}, 
\begin{equation}
F(\theta; \theta_0, \theta_1) = 
\left\{
\begin{array}{l}
1\\
\hspace{1.2 in} \mathrm{for} \; \theta \le \theta_0, \\
\cos^2[(\pi / 2) (\theta- \theta_0) / \theta_1]\\
\\
\hspace{1.2 in} \mathrm{for} \; \theta_0 \le \theta \le \theta_0  + \theta_1, \\
0 \\
\hspace{1.2 in} \mathrm{for} \; \theta \ge  \theta_0  + \theta_1,\\
\end{array}
\right.
\end{equation}
where $\theta_0$ and $\theta_1$ are parameters that define patch width. 

The bond interaction between two bonded membrane particles, $i$ and $j$, is given by~\cite{noguchi2005} 
\begin{equation}
U_{bond}(r_{ij}) = 
\left\{
\begin{array}{l}
0\\
\hspace{1.2 in} \mathrm{for} \; r_{ij} \le 1.15l_0, \\
(80 k_BT)\exp[1/(1.15l_0 - r_{ij})] / (1.33l_0 - r_{ij})\\
\\
\hspace{1.2 in} \mathrm{for} \;  1.15l_0 < r_{ij}  < 1.33l_0, \\
\infty \\
\hspace{1.2 in} \mathrm{for} \; r_{ij} \ge 1.33l_0, \\
\end{array}
\right.
\end{equation}
with $r_{ij} = |\mathbf{r}_{ij} | =  | \mathbf{r}_j - \mathbf{r}_i | $, where $\mathbf{r}_{i}$  is position of particle $i$. Additionally, an excluded volume potential is applied between all pairs of membrane particles
\begin{equation}
U_{EV}(r_{ij}) = 
\left\{
\begin{array}{l}
\infty \\
\hspace{1.2 in} \mathrm{for} \; r_{ij} \le 0.67l_0, \\
(80 k_BT)\exp[1/(r_{ij} - 0.85l_0)] / (r_{ij} - 0.67l_0)\\
\\
\hspace{1.2 in} \mathrm{for} \; 0.67l_0 < r_{ij} < 0.85l_0,\\
0\\
\hspace{1.2 in} \mathrm{for} \; r_{ij} \ge 0.85l_0.\\
\end{array}
\right.
\end{equation}
These potentials set minimum distance between membrane particles to $0.67l_0$ and the maximum bond length to $1.33l_0$. The total area, $A$, of the membrane is constrained with a potential,
\begin{equation}
U_{area} = (k_BT)(A - A_0)^2,
\end{equation}
where $A_0 = (\sqrt{3}/4) l_0^2 N_{tri}$ and $N_{tri}$, the number of triangles in the membrane surface, may vary. Membrane particles forming the edge of the surface are confined to a frame region, located a distance $r_{frame}$ into the simulation box. Within a volume of cross-section $l_0 \times 4l_0$, where the larger extension is out of the plane in which the membrane would be extended in a stretched configuration, confined membrane particles experience a flat potential of $E_{frame}$. $E_{frame}$ may be used to control the average of $r_{frame}$ and is set by comparison with tensionless simulations performed with box rescaling~\cite{matthews2013}. When confined membrane particles move out of the central part of the frame they experience a potential essentially identical to that used for excluded volume, 
\begin{equation}
U_{confine}(r) = 
\left\{
\begin{array}{l}
E_{frame} + (80 k_BT)\exp[-1/r] / (0.18l_0 - r)\\
\\
\hspace{1.2 in} \mathrm{for} \;  0 < r  < 0.18l_0, \\
\infty \\
\hspace{1.2 in} \mathrm{for} \; r \ge 0.18l_0, \\
\end{array}
\right.
\end{equation}
where $r$ is the distance of the confined membrane particle from the closest point within the flat-potential region.



\providecommand*\mcitethebibliography{\thebibliography}
\csname @ifundefined\endcsname{endmcitethebibliography}
  {\let\endmcitethebibliography\endthebibliography}{}

\end{document}